\documentclass[a4paper,12pt]{article}
\pdfoutput=1
\usepackage{epsfig}
\usepackage{amssymb}
\usepackage{amsfonts}
\usepackage{amsmath}
\usepackage{euscript}
\usepackage{verbatim}
\usepackage{latexsym}
\usepackage{graphicx}
\usepackage{caption}
\usepackage{breqn}
\usepackage{float}
\usepackage{subcaption}
\usepackage{placeins}

\usepackage{listings}

\newif\ifdtup

\catcode`\@=11

\@addtoreset{equation}{section}

\def\@normalsize{\@setsize\normalsize{15pt}\xiipt\@xiipt
\abovedisplayskip 14pt plus3pt minus3pt%
\belowdisplayskip \abovedisplayskip
\abovedisplayshortskip \z@ plus3pt%
\belowdisplayshortskip 7pt plus3.5pt minus0pt}

\def\small{\@setsize\small{13.6pt}\xipt\@xipt
\abovedisplayskip 13pt plus3pt minus3pt%
\belowdisplayskip \abovedisplayskip
\abovedisplayshortskip \z@ plus3pt%
\belowdisplayshortskip 7pt plus3.5pt minus0pt
\def\@listi{\parsep 4.5pt plus 2pt minus 1pt
     \itemsep \parsep
     \topsep 9pt plus 3pt minus 3pt}}

\relax

\catcode`@=12

\catcode`\@=11

\def\section{\@startsection{section}{1}{\z@}{3.5ex plus 1ex minus
   .2ex}{2.3ex plus .2ex}{\large\bf}}

\def\SymBoxes#1#2#3#4{\newdimen\un@t \un@t#3%
\raisebox{#1}{\rule{#2\un@t}{#4}\hskip-#2\un@t
\@tempdimb\un@t \advance\@tempdimb by-#4\@tempcntb#2\relax%
\@whilenum{\@tempcntb>0}\do{
\rule{#4}{\un@t}\hskip\@tempdimb \advance\@tempcntb by\m@ne}%
\hskip-#2\un@t \rule[\un@t]{#2\un@t}{#4}%
\rule[\un@t]{#4}{#4}\hskip-#4
\rule{#4}{\un@t}}\hskip-#4}                

\begin{document}

\newcommand{\beq}{\begin{equation}}
\newcommand{\eeq}{\end{equation}}
\newcommand{\bea}{\begin{eqnarray}}
\newcommand{\eea}{\end{eqnarray}}
\newcommand{\beas}{\begin{eqnarray*}}
\newcommand{\eeas}{\end{eqnarray*}}
\newcommand{\defi}{\stackrel{\rm def}{=}}
\newcommand{\non}{\nonumber}
\newcommand{\bquo}{\begin{quote}}
\newcommand{\enqu}{\end{quote}}
\renewcommand{\(}{\begin{equation}}
\renewcommand{\)}{\end{equation}}
\def \eqn#1#2{\begin{equation}#2\label{#1}\end{equation}}

\def\e{\epsilon}
\def\IZ{{\mathbb Z}}
\def\IR{{\mathbb R}}
\def\IC{{\mathbb C}}
\def\IQ{{\mathbb Q}}
\def\de{\partial}
\def\Tr{ \hbox{\rm Tr}}
\def\H{ \hbox{\rm H}}
\def\HE{ \hbox{$\rm H^{even}$}}
\def\HO{ \hbox{$\rm H^{odd}$}}
\def\K{ \hbox{\rm K}}
\def\Im{ \hbox{\rm Im}}
\def\Ker{ \hbox{\rm Ker}}
\def\const{\hbox {\rm const.}}
\def\o{\over}
\def\im{\hbox{\rm Im}}
\def\re{\hbox{\rm Re}}
\def\bra{\langle}\def\ket{\rangle}
\def\Arg{\hbox {\rm Arg}}
\def\Re{\hbox {\rm Re}}
\def\Im{\hbox {\rm Im}}
\def\exo{\hbox {\rm exp}}
\def\diag{\hbox{\rm diag}}
\def\longvert{{\rule[-2mm]{0.1mm}{7mm}}\,}
\def\a{\alpha}
\def\dag{{}^{\dagger}}
\def\tq{{\widetilde q}}
\def\p{{}^{\prime}}
\def\W{W}
\def\N{{\cal N}}
\def\hsp{,\hspace{.7cm}}

\def\br{\nonumber}
\def\IZ{{\mathbb Z}}
\def\IR{{\mathbb R}}
\def\IC{{\mathbb C}}
\def\IQ{{\mathbb Q}}
\def\IP{{\mathbb P}}
\def \eqn#1#2{\begin{equation}#2\label{#1}\end{equation}}

\newcommand{\C}{\ensuremath{\mathbb C}}
\newcommand{\Z}{\ensuremath{\mathbb Z}}
\newcommand{\R}{\ensuremath{\mathbb R}}
\newcommand{\rp}{\ensuremath{\mathbb {RP}}}
\newcommand{\cp}{\ensuremath{\mathbb {CP}}}
\newcommand{\vac}{\ensuremath{|0\rangle}}
\newcommand{\vact}{\ensuremath{|00\rangle}                    }
\newcommand{\oc}{\ensuremath{\overline{c}}}
\newcommand{\psizero}{\psi_{0}}
\newcommand{\phizero}{\phi_{0}}
\newcommand{\hzero}{h_{0}}
\newcommand{\psiin}{\psi_{\rh}}
\newcommand{\phiin}{\phi_{\rh}}
\newcommand{\hin}{h_{\rh}}
\newcommand{\rh}{r_{h}}
\newcommand{\rb}{r_{b}}
\newcommand{\psibnd}{\psi_{0}^{b}}
\newcommand{\psibndp}{\psi_{1}^{b}}
\newcommand{\phibnd}{\phi_{0}^{b}}
\newcommand{\phibndp}{\phi_{1}^{b}}
\newcommand{\gbnd}{g_{0}^{b}}
\newcommand{\hbnd}{h_{0}^{b}}
\newcommand{\zh}{z_{h}}
\newcommand{\zb}{z_{b}}
\newcommand{\man}{\mathcal{M}}
\newcommand{\hbr}{\bar{h}}
\newcommand{\tbr}{\bar{t}}

\begin{titlepage}
\begin{flushright}
CHEP XXXXX
\end{flushright}
\bigskip
\def\thefootnote{\fnsymbol{footnote}}

\begin{center}
{
{\bf {\large Synthetic Fuzzballs:} \\ 
\vspace{0.3cm}
{\large A Linear Ramp from Black Hole Normal Modes}
}
}
\end{center}

\bigskip
\begin{center}
Suman DAS$^a$\footnote{\texttt{suman.das@saha.ac.in}}, 
Chethan KRISHNAN$^b$\footnote{\texttt{chethan.krishnan@gmail.com}}, A. Preetham KUMAR$^b$\footnote{\texttt{adepu@iisc.ac.in}}, Arnab KUNDU$^a$\footnote{\texttt{arnab.kundu@saha.ac.in}}

\end{center}

\renewcommand{\thefootnote}{\arabic{footnote}}

\begin{center}

$^a$ {Theory Division, Saha Institute of Nuclear Physics,\\
A CI of Homi Bhabha National Institute,\\
1/AF, Bidhannagar, Kolkata 700064, India}\\

\vspace{0.2cm}

$^b$ {Center for High Energy Physics,\\
Indian Institute of Science, Bangalore 560012, India}\\

\end{center}

\noindent
\begin{center} {\bf Abstract} \end{center}
We consider a black hole with a stretched horizon as a toy model for a fuzzball microstate. The stretched horizon provides a cut-off, and therefore one can determine the normal (as opposed to quasi-normal) modes of a probe scalar in this geometry. For the BTZ black hole, we compute these as a function of the level $n$ and the angular quantum number $J$. Conventional level repulsion is absent in this system, and yet we find that the Spectral Form Factor (SFF) shows clear evidence for a dip-ramp-plateau structure with a linear ramp of slope $\sim 1$ on a log-log plot, with or without ensemble averaging.  We show that this is a robust feature of stretched horizons by repeating our calculations on the Rindler wedge (times a compact space).  We also observe that this is {\em not} a generic feature of integrable systems, as illustrated  by standard examples like integrable billiards and random 2-site coupled SYK model, among others. The origins of the ramp can be traced to the hierarchically weaker dependence of the normal mode spectrum on the quantum numbers of the compact directions, and the resulting quasi-degeneracy. We conclude by noting an analogy  between the 4-site coupled SYK model and the quartic coupling responsible for the non-linear instability of capped geometries. Based on this, we speculate that incorporating probe self-interactions will lead to stronger connections to random matrix behavior.

\vspace{1.6 cm}
\vfill

\end{titlepage}

\setcounter{footnote}{0}

\section{Introduction}

The nature of the black hole horizon and whether spacetime ends there, has been a point of debate ever since the discovery of the Schwarzschild metric \cite{Schwarzschild, Droste}. This debate has gone through various iterations, and has recently been revived in the context of the information paradox \cite{Hawking, Page, Mathur, AMPS}. In the present paper, we will view this as a question about the meaning and interpretation of black hole microstates in string theory \cite{SV, LM1}. These have the same macroscopic charges as a black hole in the bulk, but at least in the supergravity approximation, candidate microstate solutions \cite{LM1, KST, superstrata} cap off before the horizon and are completely regular. They provide the foundation for the fuzzball program \cite{Mathur, Mathur1, BenaReview, Masaki}. 

\noindent
There are a few facts about these solutions that we find remarkable:
\begin{itemize}
\item It is non-trivial that solutions with a throat that caps off before the horizon {\em exist} in low energy string theory, even though in classical general relativity they are ruled out by no hair theorems and the like.
\item Far from the throat they are locally indistinguishable from the black hole -- they have the same charges as the black hole, but the ``hair'' exists only in the cap region deep in the throat. In other words, classically these solutions are black hole mimickers all the way up to the horizon, and in that sense are quite distinct from say stars or even neutron stars. This qualitative feature would be surprising, if these solutions had no significance for black holes.
\item There are whole moduli spaces of such solutions. This is necessary if they are to be interpreted as microstates of the black hole (in the classical limit).
\item  In some limited cases\footnote{Specifically, in the case of the 2-charge black hole, where one needs $\alpha'$-corrections to have a finite size horizon.}, these spaces can be fully determined and successfully geometrically quantized to yield the entropy of the black hole \cite{Rychkov} including the precise numerical coefficient \cite{Avinash}.
\end{itemize}
\noindent
But there are also flies in the ointment:
\begin{itemize}
\item Modulo some minor caveats \cite{JMART}, microstate solutions are only known for BPS black holes which are at zero temperature. 
\item An obvious criticism regarding supergravity microstates is that the more complicated profile functions that capture generic fuzzballs will have high (presumably Planckian) curvature and are not reliable in supergravity or even tree level string theory\footnote{Since this is such a basic criticism, we suspect that the attitude of the fuzzball community to this question is that at least for BPS protected states, the SUGRA solutions are simply a place-holder for the more reliable solutions that presumably must also exist in the full string theory.}. But once one is in the regime of quantum string theory, there is an operational lack of clarity (at least in our opinion) about what it means to say that spacetime caps of at the horizon for individual microstates.
\item It has been argued \cite{Sen} that in the 2-charge system, the fuzzball solutions should be {\em added} to the entropy of the black hole, depending on the duality frame\footnote{Arguments against this suggestion can be found in \cite{MathurReply}.}.
\item In the 3-charge case, the solution spaces discovered so far (``superstrata'' \cite{superstrata}) only account for a subleading fraction of the black hole entropy. There are reasons to suspect that they may not all be accessible in supergravity \cite{PureHiggs}.
\item Any claim that microstates cap off at the horizon raises various dynamical/thermodynamic questions, which have relatively simple (or at least well-known) explanations in terms of the black hole picture, but are very challenging in terms of microstates. The most basic of these are questions of smooth infall, the nature of Kruskal coordinates and the interpretation of Hawking's original computation of Hawking temperature \cite{particle}. 

\item More generally, the problem of understanding various aspects of horizon physics in a dynamical setting via an ensemble of horizonless microstates, is clearly an outstanding challenge.
\end{itemize}

In this paper, we wish to discuss some aspects of the last bullet point. Specifically, we want to explore the connections between fuzzball microstates, and developments on scrambling and chaos that have appeared regarding black hole horizons in recent years \cite{Sekino, Butterfly, MSS}. These results are generic when the black hole is at finite temperature, so one can try to identify crude calculations motivated by the fuzzball program, that can reproduce some of these features.

One heuristic expectation \cite{cotler} from the recent results on scrambling and chaos at the horizon is that the spectrum of black hole microstates should exhibit behavior familiar from random matrix theory (RMT) -- with level repulsion in eigenvalues, and with level spacing that is exponentially small $\sim \exp(-1/G)$. As a corollary, we would expect the Spectral Form Factor (SFF) \cite{cotler} to exhibit a dip-ramp-plateau structure\footnote{See \cite{SFF} for a very incomplete sampling of papers that investigate the SFF in the wake of \cite{cotler}.} with a linear ramp and a plateau, instead of the exponential decay at late times that is the signature of the information paradox \cite{eternal, Barbon}. Evading information paradox in the sense of \cite{eternal, Barbon} is tautological in the fuzzball paradigm, because the absence of a horizon (by construction) prevents the exponential decay  -- the cap would result in the decay being replaced by a plateau. But the more ``difficult'' feature to see from the fuzzball spectrum is the level repulsion/ramp implicit in RMT\footnote{The crude expectation is that the ramp in the SFF is a manifestation of level repulsion. But precise statements are hard to come by, and we will discuss various caveats. In the present paper, one of our points is that we will find scenarios where the ramp emerges despite the absence of conventional level repulsion or ensemble averaging. This is not quite a contradiction, as we will elaborate in a concluding subsection. We thank Julian Sonner for discussions on related matters.}.  


Reproducing these RMT expectations directly, even in the dual CFT, is a tall order. This is because we need to know the spectrum of a theory, as complicated as large-$N$ $\mathcal{N}=4$ SYM on a sphere, at energies that scale with positive powers of $N$. This is the regime where the spectrum is supposed to capture the microstates of a dual large AdS black hole. The hope is that examples like SYK \cite{Polchinski} retain aspects of this spectrum in tractable toy models. Doing a calculation from the bulk side that can see (some of) this physics is of course significantly harder. To identify the spectrum exactly (and in particular to see its discreteness and exponential level spacing) one would have to solve string theory in the relevant asymptotically AdS geometry. In the fuzzball program, what one instead has access to are merely solution spaces of supergravity which serve as a proxy for the quasi-degenerate microstates of a black hole with some given macroscopic charges (including mass). There is a further complication related to the fact that in order to study chaos and RMT behavior, one needs the fuzzball microstates of a finite temperature black hole. Explicit supergravity solution spaces \cite{LM1, superstrata} are meaningful only in the zero temperature BPS case. Given this very incomplete state of affairs, part of the challenge is to come up with a {\em tractable} question that one can ask, whose answer may yield hints or suggestions of RMT behavior from the bulk side. 

In this paper, we will make the most rudimentary proposal for such a calculation -- we will use the normal mode spectrum of a  black hole with a stretched horizon as a probe of potential RMT-like behavior. The results we are aiming for are expected to be generic at finite temperature, so we may be able to learn something despite the absence of controllable solutions,  by explicitly introducing the stretched horizon {\em by hand}.

For a scalar field in a black hole background, one typically computes {\em quasi-normal} modes, which correspond to infalling boundary conditions at the horizon. These are complex modes, and  correspond to decaying waves as opposed to standing waves. By introducing a stretched horizon on the other hand, it becomes possible to define and compute {\em normal} modes which arise from Dirichlet boundary conditions at the stretched horizon. For a {\em free} scalar field in a geometry with a ``cap'' (in this case, the stretched horizon), a reasonable expectation is that the level spacing distribution of these modes would be Poisson (or perhaps roughly equally spaced). See \cite{Martinec} for some general comments about the level spacing distributions on capped geometries.  

We will find that even though this expectation is true, there are some intriguing wrinkles in the statement due to the spectrum of normal modes being not just a function of the level $n$, but also\footnote{We will consider stretched horizons on the BTZ black hole and the (Rindler wedge)$\times$(compact space) in this paper. Radial parts of wave equations in higher dimensional black holes give rise to less tractable special functions, but we do not expect qualitative differences.} of the angular quantum number $J$. Remarkably, despite the absence of conventional level repulsion, we find that the SFF shows the characteristic dip-ramp-plateau (DRP) structure with a clear linear ramp. This is true with or without ensemble averages\footnote{A natural ensemble average one can do in our problem is over a small Gaussian distributed window of stretched horizon radii.}. The presence of a ramp is usually viewed as being associated to RMT behavior\footnote{As already noted, statements of this type are fairly heuristic. We will present more detail, later.}, even though in some integrable cases, a semblance of a ramp may emerge {\em after} an ensemble average  when there are random couplings. One example of this latter type we are aware of is the 2-site coupled SYK model \cite{Jeff}. But even in this case, we will show that the slope of the ramp is {\em not} consistent with $\sim 1$. In our case on the other hand, there is nothing random once the stretched horizon is chosen, and we can see the linear ramp of slope $\sim 1$ with or without any averaging. To emphasize that this is not a generic feature of Poisson distributed systems, we explicitly check that the SFF has no ramp for various well-known examples of integrable systems with Poisson statistics. 

We can get a better understanding of the origin of the ramp by considering the normal modes $\omega(n,J)$ restricted over specific ranges of $n$'s and $J$'s, and studying the changes in the SFF. In particular, we find that the ramp is present in the SFF as long as we are summing over a sufficiently large range of $J$'s irrespective of the range of $n$ that we retain. In particular, just retaining one value of $n$ while summing over a big enough range of $J$'s is sufficient to produce the ramp. On the contrary, if we are working with too few $J$'s the ramp goes away irrespective of the range of $n$ that we retain. These results are a clear indication that the $J$-dependence of the spectrum is responsible for the ramp seen in the SFF. Our primary interest will be in the behavior of the SFF due to the $J$-direction, and we have checked that our qualitative results are stable as we change the $J$-cut-off.
 
\begin{figure}[h]
\begin{subfigure}{0.47\textwidth}
    \centering
    \includegraphics[width=\textwidth]{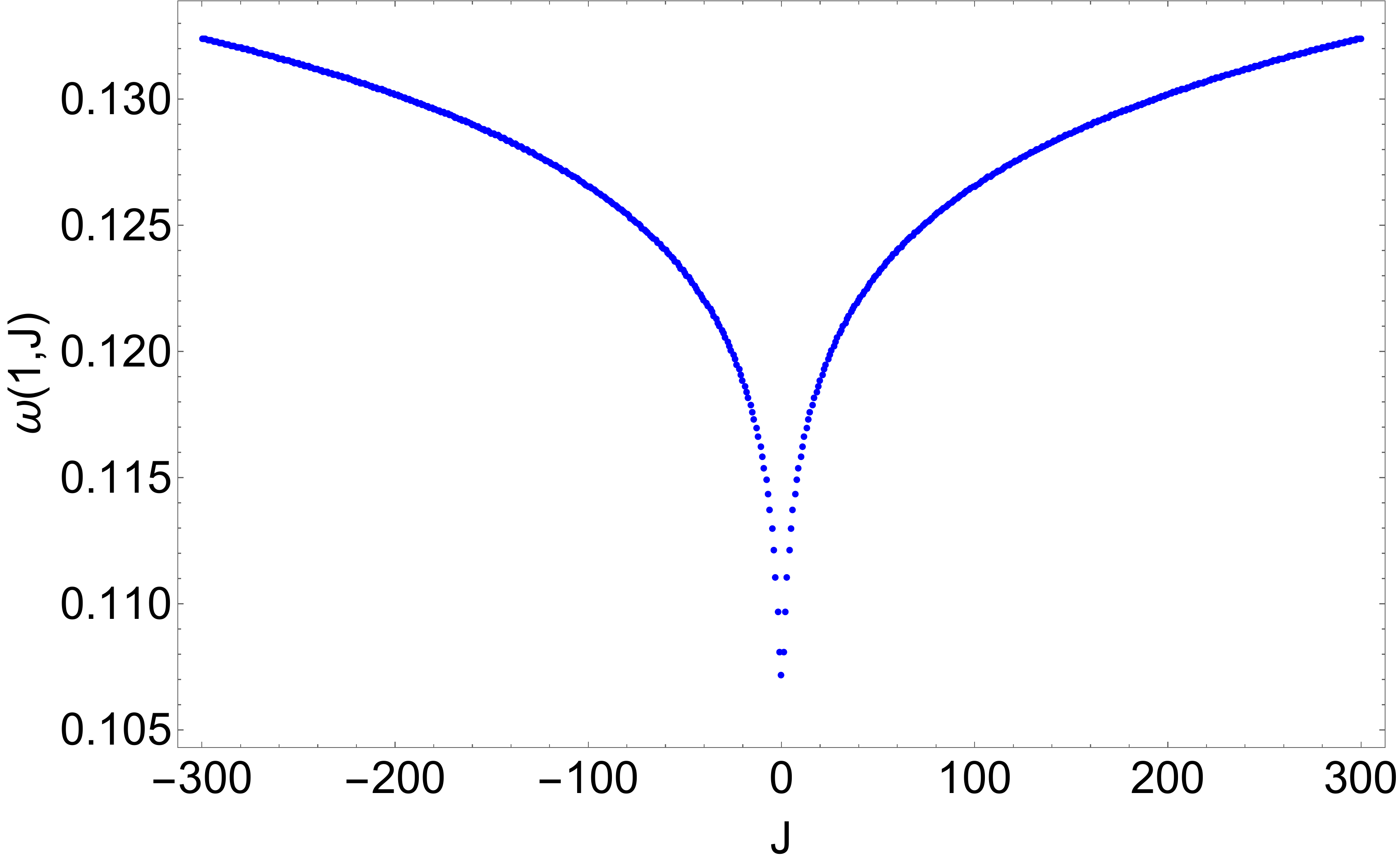}
    \end{subfigure}
    \hfill
    \begin{subfigure}{0.47\textwidth}
    \includegraphics[width=\textwidth]{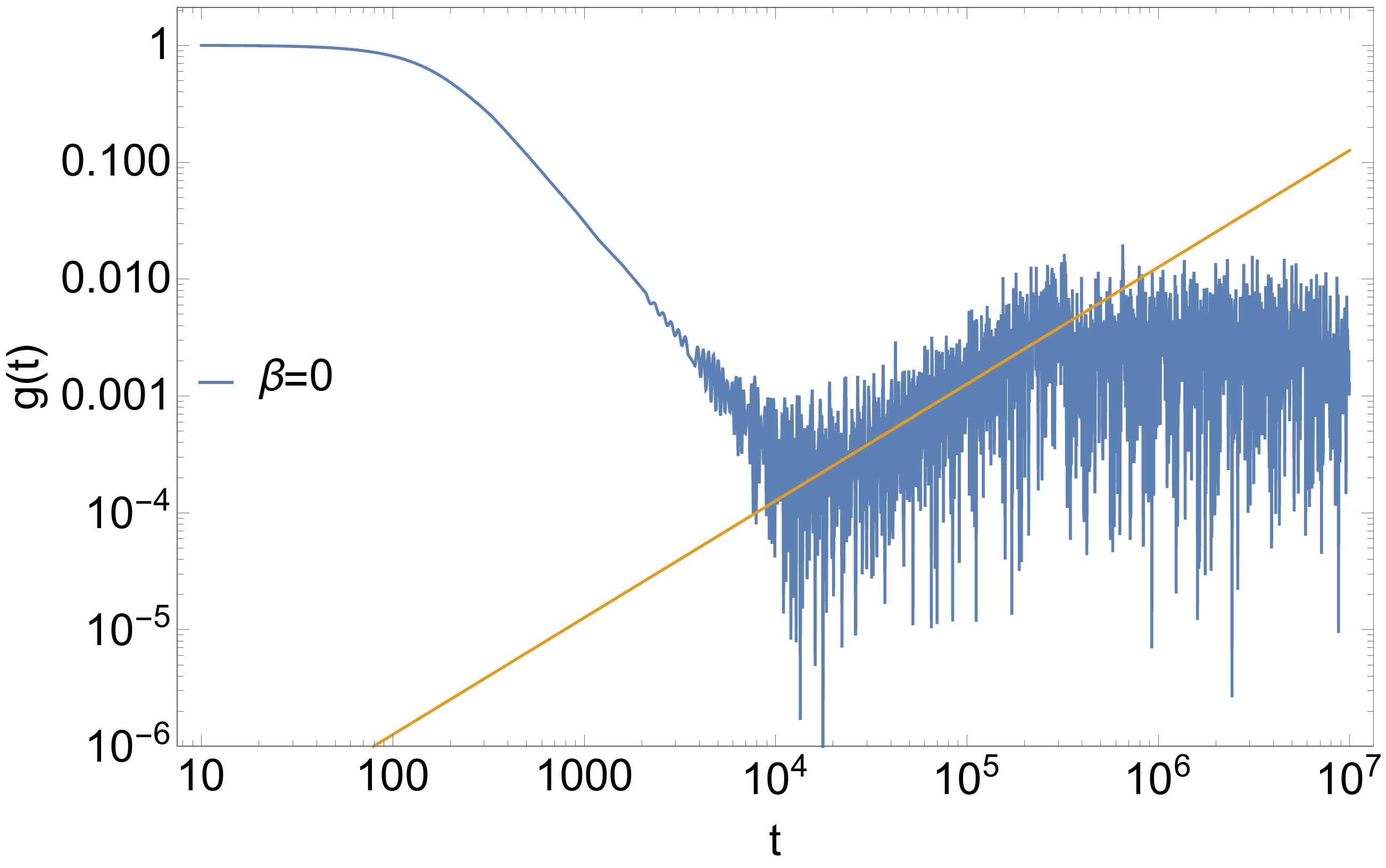}
    \end{subfigure}
    \caption{Plot of energy levels $\omega(1,J)$ (left) and the $\beta=0$ spectral form factor (right) for BTZ black hole normal modes with $z_0=30$, $n=1$ and $J_{cut}=300$. The notations are defined in later sections, but the general message should be clear.}
    \label{btz_level}
\end{figure}

\begin{figure}[H]
\begin{subfigure}{0.46\textwidth}
    \centering
    \includegraphics[width=\textwidth]{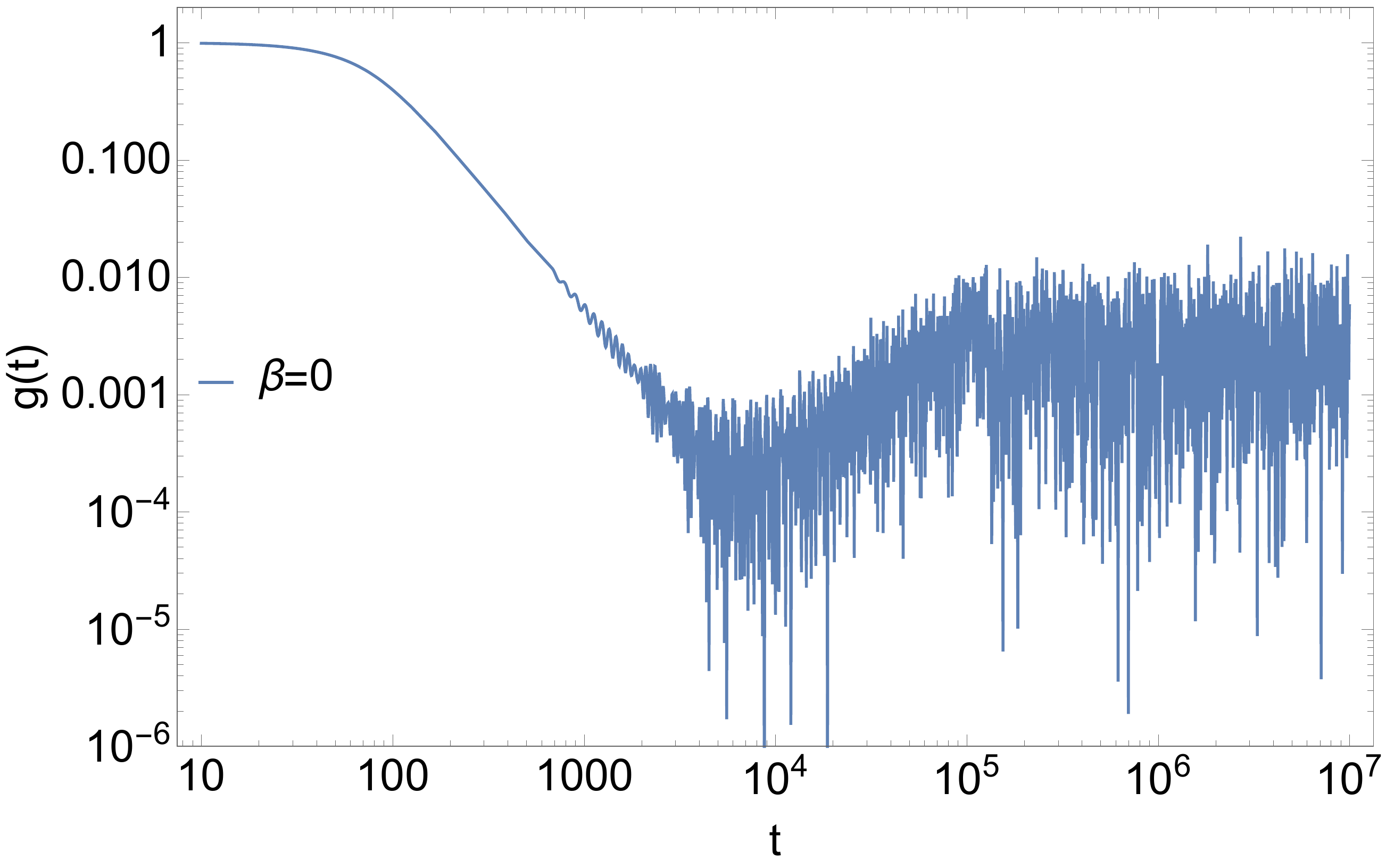}
    \end{subfigure}
    \hfill
    \begin{subfigure}{0.46\textwidth}
    \includegraphics[width=\textwidth]{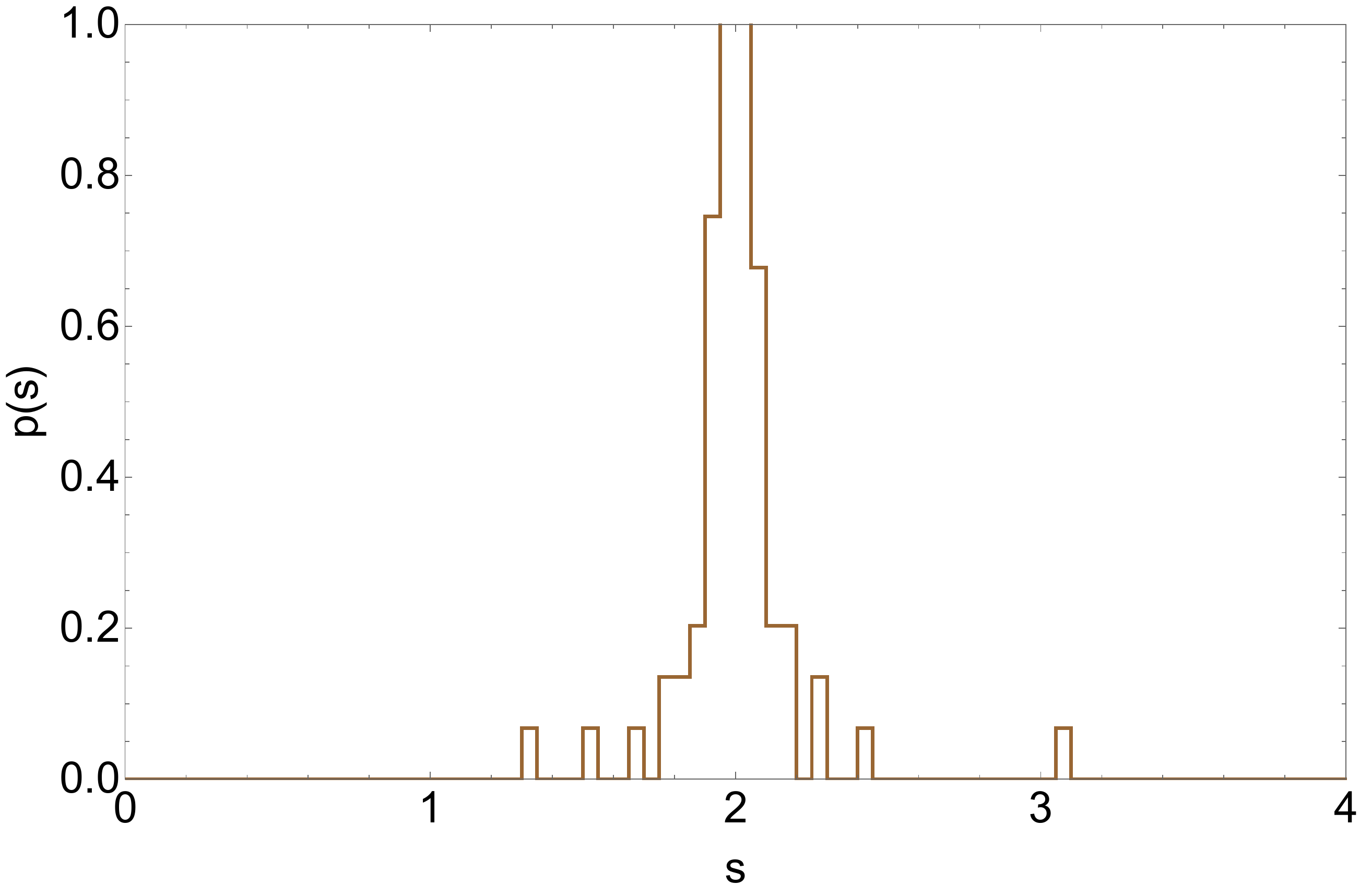}
    \end{subfigure}
    \caption{SFF (left) and its corresponding level spacing plot (right) for $3d$ Rindler normal modes with $\xi_0=20$, $J_{cut}=300$, $n$ is fixed to one and $\beta=0$. Other parameters $R$ and $a$ are fixed to  $2$ and $1$. The bin size on the right is 0.05.}
    \label{sff_rindler_j}
\end{figure}

To further clarify the $n$ and $J$ dependence of the spectrum, we study the level spacing distributions of $\omega(n,J)$ for fixed $n$ or fixed $J$. For fixed $J$, the levels are roughly equally spaced and qualitatively similar to those of the harmonic oscillator. This is consistent with the observation in the previous paragraph that there is no ramp in the SFF when we work with a small enough range of $J$'s. The level spacing distribution of $\omega(n,J)$ for fixed $n$ (ie., as a function of $J$) on the other hand, has non-trivial structure. It is not equally spaced, as is evident from the actual plot of the levels, see Figure \ref{btz_level} left panel. But it is also decidedly not Poisson as is clear from the level-spacing distribution, see Figure \ref{btz_j} and \ref{sff_rindler_j} right panel. If we view this level spacing as a ``generalized'' RMT distribution\footnote{The structure is loosely consistent with a distribution of the form $P(s) \sim  s^n \exp(-\alpha \ s^2)$ 
with large values of $n$. More elaborate numerical work is required to conclusively extract the form and/or fix the coefficients.}, then it is tempting to view the ramp as being  a result of this ``generalized'' RMT behavior in the $J$-direction. We leave a detailed analysis of this issue for future work. Here, we will limit ourselves to observing that the level spacing in the $J$-direction is non-trivial and that this is correlated with the presence of the ramp.
%

As we steadily increase the cut-off in $n$, the peaks of the level spacing distribution are effectively submerged and we end up finding something resembling a Poisson distribution, see eg. Figure \ref{btz_nj}. Despite this, we see that the ramp remains intact -- clearly the SFF is sensitive to the underlying ``generalized'' RMT behavior in the $J$-direction, even when the cut-off in $n$ is fairly large.


The paper is organized as follows. In the upcoming section, we introduce a stretched horizon in a BTZ black hole geometry and  compute the normal modes. A similar calculation is later also done for a stretched horizon on a (Rindler wedge) $\times$ (a circle)\footnote{The ramp structure is unaffected if we replace the circle with another compact space.}. In both cases, using these normal modes, we can compute the SFF and we see the linear ramp. We also see that the level spacing distribution does not have conventional level repulsion, as we outlined above\footnote{Various technical aspects are also discussed, but we will not emphasize them in the Introduction.}. We also study various well-known Poisson distributed integrable systems and show that they do {\em not} have a ramp. 

In a concluding section, we make some observations regarding what it might take to see more conclusive evidence for fuzzball RMT behavior, in calculations involving probes on the stretched horizon. We discuss the utility of studying a self-interacting probe scalar (or fermion) instead of a free probe, on the stretched horizon. We make connections with higher-$q$ SYK models, as well as observations on the the non-linear instability of capped geometries \cite{Reall}. We also discuss the expected non-perturbative smallness of level spacing -- we present the double-well potential as an example to illustrate that interactions are likely key, here as well. In particular, interactions have the ability to produce level spacings that are non-perturbatively small in the coupling. A final point we emphasize is the role of compact dimensions in our results -- the quantum numbers in the compact dimensions played a crucial role in both the BTZ case as well as the Rindler case in producing the ramp. The reader who is willing to believe our numerical calculations in the intermediate sections may want to skip ahead to this final section, which may be of some general interest to the fuzzball program.

\section{Case Study I: Scalar Fields on the BTZ black Hole}


In this section we will solve the free scalar field equation on the BTZ black hole background, see eg. \cite{Keski-Vakkuri}, but with Dirichlet boundary condition at some finite $r=r_{0} > r_h$ instead of the usual infalling boundary condition at the horizon $r=r_h$. The latter leads to quasi-normal boundary conditions (see eg., \cite{Horowitz}) whereas we are interested in computing normal modes of the stretched horizon. Our calculations are also closely related to the higher dimensional calculations of \cite{Festuccia} and their BTZ analogue \cite{ChenYang}\footnote{We have fixed some errors in \cite{ChenYang}.}.

Consider a scalar field $\Phi$ of mass $m$ in the BTZ background given by the metric,
\begin{equation}\label{BTZ_metric}
    ds^2=-(r^2-r_h^2)dt^2+\frac{dr^2}{(r^2-r_h^2)}+r^2 d\psi^2
\end{equation}
where $-\infty<t<\infty$, $0<r<\infty$ and $0\le \psi< 2\pi$. Equation of motion 
\begin{equation}\label{eom1}
    \Box \Phi=m^2 \Phi
\end{equation}
with $\Box \Phi \equiv \frac{1}{\sqrt{|g|}}\partial_{\mu}\left(\sqrt{|g|}\partial^{\mu}\Phi\right)$ 
can be solved by exploiting the isometries in the $t$ and $\psi$ direction by writing 
\begin{equation}
    \Phi(t,r,\psi)=\sum_{J,\omega}e^{i J \psi} e^{-i \omega t} \frac{\phi_{\omega,J}(r)}{\sqrt{r}}
\end{equation}
where $J$ is integer (due to periodicity in $\psi$ direction). In the following, we suppress the $\omega,J$ subscript of $\phi$. With the above form of $\Phi$, the radial part of \eqref{eom1} satisfies,
\begin{equation}
    (r^2-1)^2\frac{d^2\phi(r)}{dr^2}+2r(r^2-1)\frac{d\phi(r)}{dr}+\omega^2\phi(r)-V(r)\phi(r)=0 \label{radial_eom}
\end{equation}
where \begin{equation*}
    V(r)=(r^2-1)\left[\frac{1}{r^2}\left(J^2+\frac{1}{4}\right)+\nu^2-\frac{1}{4} \right], \hspace{2cm} \nu^2=1+m^2.
\end{equation*}
We have set $r_h=1$ here and in the rest of the calculations. 
Solution\footnote{We will be concerned with the massless scalar field in this paper. In this case, the $J=0$ case requires special treatment of the solution, and can be expressed in terms of Meijer G functions. But it turns out that the $J=0$ mode does not contribute in our later computations, because demanding normalizable fall-offs at the boundary \eqref{falloff} removes them.} of equation \eqref{radial_eom} is given in terms of hypergeometric functions as
\begin{equation}\label{sol1}
    \phi(r)=\sqrt{r}(1-r^2)^{-\frac{i \omega}{2}}\left(C_1 r^{-i J} G(r)+C_2 r^{i J} H(r) \right),
\end{equation}
where,
\begin{align*}
    G(r)&={}_2F_1\left(\frac{1}{2}\left( -i \omega-i J-\nu+1\right),\frac{1}{2}\left(-i\omega-i J+\nu+1 \right);1-i J,r^2\right) \\
    H(r)&={}_2F_1\left(\frac{1}{2}\left( -i \omega+i J-\nu+1\right),\frac{1}{2}\left(-i\omega+i J+\nu+1 \right);1+i J,r^2\right).
\end{align*}
For convenience let us also introduce a different radial coordinate $z$  (``tortoise coordinate'') which is defined as,
\begin{equation}\label{ztor}
    z(r)=\frac{1}{2}\ln \frac{r+1}{r-1},
\end{equation}
for $1<r<\infty$. In this coordinate $z\rightarrow\infty$ is the horizon and $z\rightarrow0$ is the AdS boundary. This allows us to choose the normalizable mode conveniently at the boundary. Together then, the two defining conditions for our BTZ normal modes are
\begin{equation}
    \phi(z)\rightarrow \begin{cases}
    C(\omega,J)z^{\frac{1}{2}+\nu}, & \text{as $ z\rightarrow0$} \\
    0, & \text{as $ z\rightarrow z_0$},
    \end{cases} \label{falloff}
\end{equation}
for some finite $z=z_0$, which is the location of the stretched horizon in tortoise coordinates. The precise coefficient $C(\omega,J)$ can be determined, and we will discuss it momentarily.

\subsection{Stretched Horizon Normal Modes}

Our primary motivation for working with the BTZ black hole (instead of say Schwarzschild) is that the radial equation is solvable in terms of hypergeometric functions. In higher dimensions, typically these equations are Heun equations which are less tractable. This simplification enables us to do our normal mode calculations semi-analytically, which should be contrasted to higher dimensions where one has to resort to a more full-fledged numerical approach (see eg. the quasi-normal mode calculations in \cite{Horowitz}).

Near the boundary ($z\rightarrow0$), equation \eqref{sol1} can be approximated as
%
%
\begin{align}\label{sol3}
    \phi_{bdry}(z)= & \left((-1)^{-\frac{1}{2}} z^{\frac{1}{2}-\nu}\left( (-1)^{\frac{i J+\nu}{2}}  \gamma(J, \nu) C_1+(-1)^{\frac{-i J+\nu}{2}}  \gamma(-J, \nu) C_2 \right)+O(z)^{\frac{3}{2}-\nu}\right)\nonumber \\
    &+\left((-1)^{-\frac{1}{2}} z^{\frac{1}{2}+\nu}\left((-1)^{\frac{i J-\nu}{2}}  \gamma(J, -\nu) C_1+(-1)^{-\frac{i J+\nu}{2}}  \gamma(-J, -\nu) C_2  \right) +O(z)^{\frac{3}{2}+\nu} \right),
\end{align}
where
\begin{align*}
    \gamma(J, \nu) &\equiv \frac{\Gamma(1-i J)\Gamma(\nu)}{\Gamma\left(\frac{1}{2}(1-i \omega-i J+\nu)\right) \Gamma\left(\frac{1}{2}(1+i \omega-i J+\nu)\right) }, 
\end{align*}
Then boundary condition at $z\rightarrow0$ sets the first term of equation \eqref{sol3} to zero. This implies
\begin{equation}\label{c1c2}
    C_2=-(-1)^{i J}\frac{\gamma(J,\nu)}{\gamma(-J,\nu)}C_1,
\end{equation}
effectively fixing the coefficient $C(\omega,J)$ from the previous subsection. We will not write  $C(\omega,J)$ explicitly, because we do not need it.

We are interested in a stretched horizon type of scenario -- we want to place the cutoff close to the horizon i.e. $z_0$ is large (in practice we find that $z_0 \gtrsim 10$ does the job), or equivalently $r_0$ is only slightly bigger than $r_h=1$. Near the horizon, by expanding the hypergeometric functions, \eqref{sol1} can be approximated as
\begin{equation}\label{sol4}
    \frac{ \phi_{hor}(r)}{\sqrt{r}}\approx \Big(C_1 \ \tilde\gamma(J,\omega)+C_2\ \tilde\gamma(-J,\omega) \Big)(1-r^2)^{-\frac{i \omega}{2}}+\Big(C_1\ \tilde\gamma(J,-\omega) +C_2\ \tilde\gamma(-J,-\omega) \Big)(1-r^2)^{\frac{i \omega}{2}},
\end{equation}
where
\begin{align*}
    \tilde\gamma(J,\omega) &\equiv \frac{\Gamma(1-i J)\Gamma(i \omega)}{\Gamma\left(\frac{1}{2}(1+i \omega-i J+\nu)\right) \Gamma\left(\frac{1}{2}(1+i \omega-i J-\nu)\right) }, 
\end{align*}
%
Using \eqref{c1c2}, we can write
\begin{equation}\label{sol5}
    \phi_{hor}(r)\approx C_1 \sqrt{r} \left(P(1-r^2)^{-\frac{i \omega}{2}}+Q e^{\pi \omega}(1-r^2)^{\frac{i\omega}{2}} \right),
\end{equation}
where
\begin{align}
    P &=\frac{\left(e^{J\pi}-e^{-J\pi}\right)\Gamma(1-i J)\Gamma(i\omega)}{\left(e^{J\pi}+e^{\pi(\omega+i \nu)}\right) \Gamma\left(\frac{1}{2}(1+i\omega-i J-\nu)\right) \Gamma\left(\frac{1}{2}(1+i\omega-i J+\nu)\right)} \label{PQ1}\\
    Q &=\frac{e^{-\pi\omega}\left(e^{J\pi}-e^{-J\pi}\right)\Gamma(1-i J)\Gamma(-i\omega)}{\left(e^{J\pi}+e^{\pi(-\omega+i \nu)}\right) \Gamma\left(\frac{1}{2}(1-i\omega-i J-\nu)\right) \Gamma\left(\frac{1}{2}(1-i\omega-i J+\nu)\right)}.\label{PQ2}
\end{align}
The boundary condition at $r=r_0$ i.e. $\phi_{hor}(r=r_0)=0$ implies
\begin{align*}
    & P(1-r_0^2)^{-\frac{i \omega}{2}}+Q e^{\pi \omega}(1-r_0^2)^{\frac{i\omega}{2}}=0\\
    \Rightarrow &(r_0^2-1)^{i \omega}=-\frac{P}{Q}.
\end{align*}
In an Appendix we show that $|P|=|Q|$. So equating arguments of the last equation we get
\begin{equation}\label{solfinal}
    \text{Arg}\left(\frac{P}{Q}\right)=\text{Arg}(-1)+\omega \log(r_0^2-1).
\end{equation}
In $z$ coordinate this relation transforms to the following,
\begin{equation}
    \text{Arg}\left(\frac{P}{Q}\right)-(\omega \log4-2\omega z_0)=\text{Arg}(-1)=\pi
\end{equation}
Explicit form of $\text{Arg}\left(\frac{P}{Q}\right)$ is computed in one of the Appendices. For fixed $z_0$ and $\nu$ this equation can be numerically solved to obtain $\omega$. Since this is a phase equation, the modes depend on a free integer $n$. We consider the massless limit (i.e. $\nu=1$) of the probe scalar field $\Phi$ and solve this equation numerically in Mathematica to get these normal modes. 

In the next section we will see how these normal modes give rise to the dip-ramp-plateau structure and how it changes with $J$,  $n$, $z_0$ etc. It is worth mentioning here that to find normal modes for cutoffs that are not too close to the horizon, we have to use equation \eqref{sol1} directly. We will only need this to illustrate that the ramp structure goes away as the stretched horizon is moved away from the horizon.  


\section{Dip, Ramp and Plateau}


The Spectral Form Factor (SFF) is defined as \cite{cotler}
\begin{equation}
    g(\beta,t)=\frac{|Z(\beta,t)|^2}{|Z(\beta,0)|^2}. \label{sff}
\end{equation}
For a given quantum mechanical system $Z(\beta,t)=\text{Tr}\left[e^{-(\beta-it)H}\right]$ where $\beta$, $t$ and $H$ are  inverse temperature, time and the Hamiltonian respectively.

Our system is a free field theory on the black hole background with a stretched horizon. We will interpret the normal modes found using the discussion of the previous section as eigenvalues of a quantum mechanical system, and define the SFF using them. The eigenvalues $\omega_{n,J}$  are labeled by $n$ and $J$. In principle $J$ can go from $-\infty$ to $\infty$ and $n$ from $1$ to $\infty$ but for our calculation we will truncate both the series at some $|J_{cut}|$ and $n_{cut}$. With this truncated series, we have 
\begin{align}
    Z(\beta,t) 
    = \sum_{\omega} e^{-(\beta-it)\omega}= \sum_{J=-J_{cut}}^{J_{cut}} \sum_{n=1}^{n_{cut}} e^{-(\beta-it)\omega_{n,J}}. \label{Z}
\end{align}
This is what we will use in our definition of the SFF. Note that strictly speaking this is a new definition -- we are extending the usual definition \eqref{sff} to a system with an $O(2)$ symmetry\footnote{\label{GSFF}It may be useful to define a Gibbsian Spectral From Factor (GSFF) or Grand Canonical Spectral Form Factor (GCSFF), via 
\bea
g_G(\beta, \mu,t,\phi) =\frac{|Z_G(\beta,\mu,t,\phi)|^2}{|Z_G(\beta,\mu,0,0)|^2},\ \ {\rm where} \ \  
Z_G(\beta, \mu,t,\phi)\equiv \text{Tr}\left[e^{-(\beta+i t)H-(\beta \ \mu+i \phi)J}\right]
\eea
in discussions of systems with symmetry.  Here, $\mu$ is a chemical potential.  We have taken the symmetry to be a rotational $O(2)$ symmetry with charge $J$ and the dual circle $\phi$ in the equations above. This is for concreteness, but obvious generalizations apply to more number of symmetries once we introduce more chemical potentials. This seems like a very natural extension of the conventional SFF to systems with global symmetries. However, to our knowledge it has not been defined or studied -- in particular, it will be interesting to see how RMT-like behavior is encoded in such an object. The SFF we are studying in this paper can be viewed as a special case, $g(\beta,t)=g_G(\beta,0,t,0)$, which already contains interesting information.}. We will see that this still retains various pieces of interest, including (perhaps surprisingly) the ramp. We have also checked that the qualitative behaviors that we will be interested in, remain intact as we slide the cut-offs, for large enough cut-offs.



\subsection{SFF of BTZ Black hole}

\begin{figure}[h]
    \centering
    \includegraphics[width=.6\textwidth]{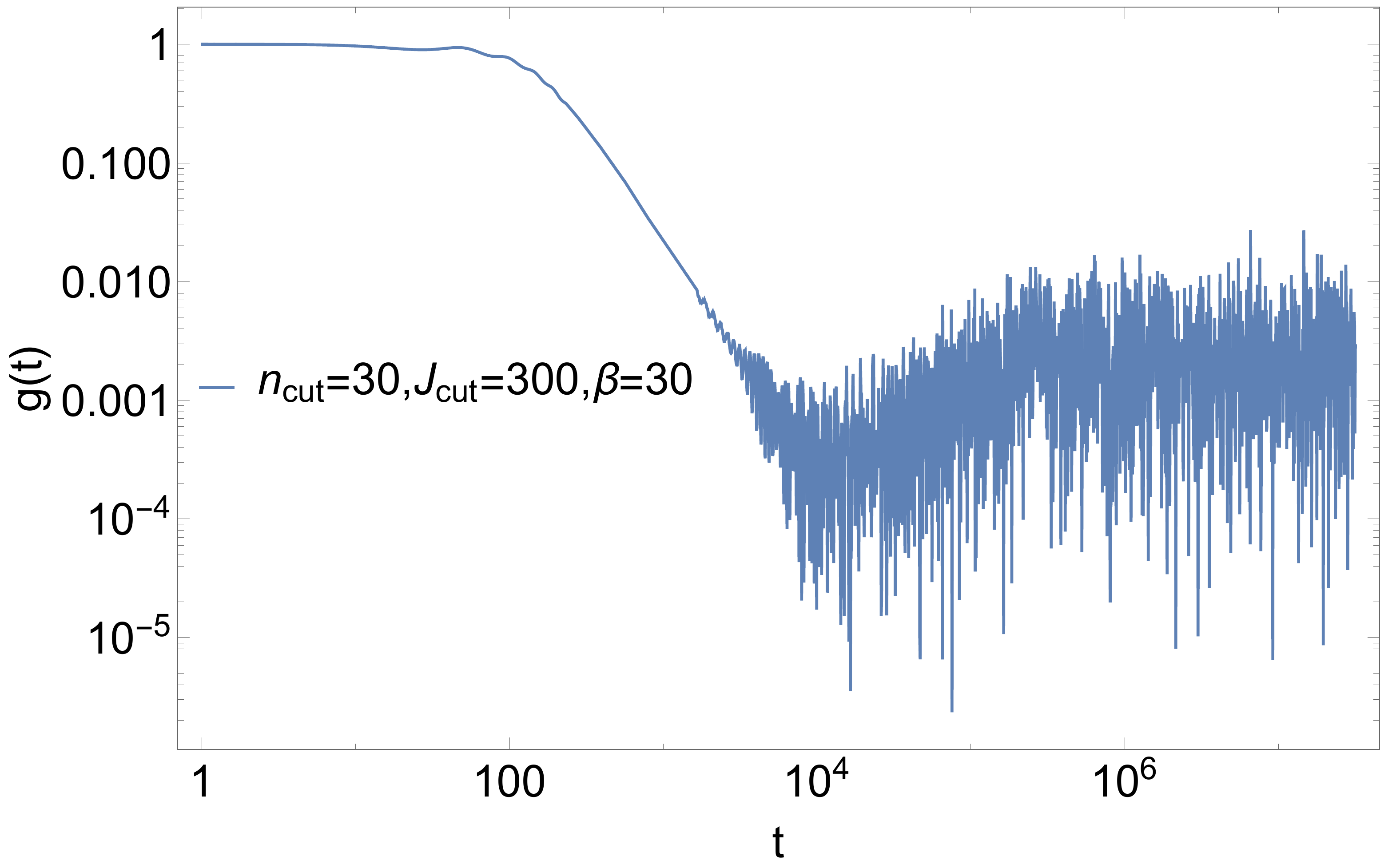}
   \caption{SFF for BTZ black hole normal modes with $z_0=30$, $J_{cut}=300$, $n_{cut}=30$ with $\beta=30$. }
    \label{btz_nj1}
\end{figure}

\begin{figure}[H]
\begin{subfigure}{0.48\textwidth}
    \centering
    \includegraphics[width=\textwidth]{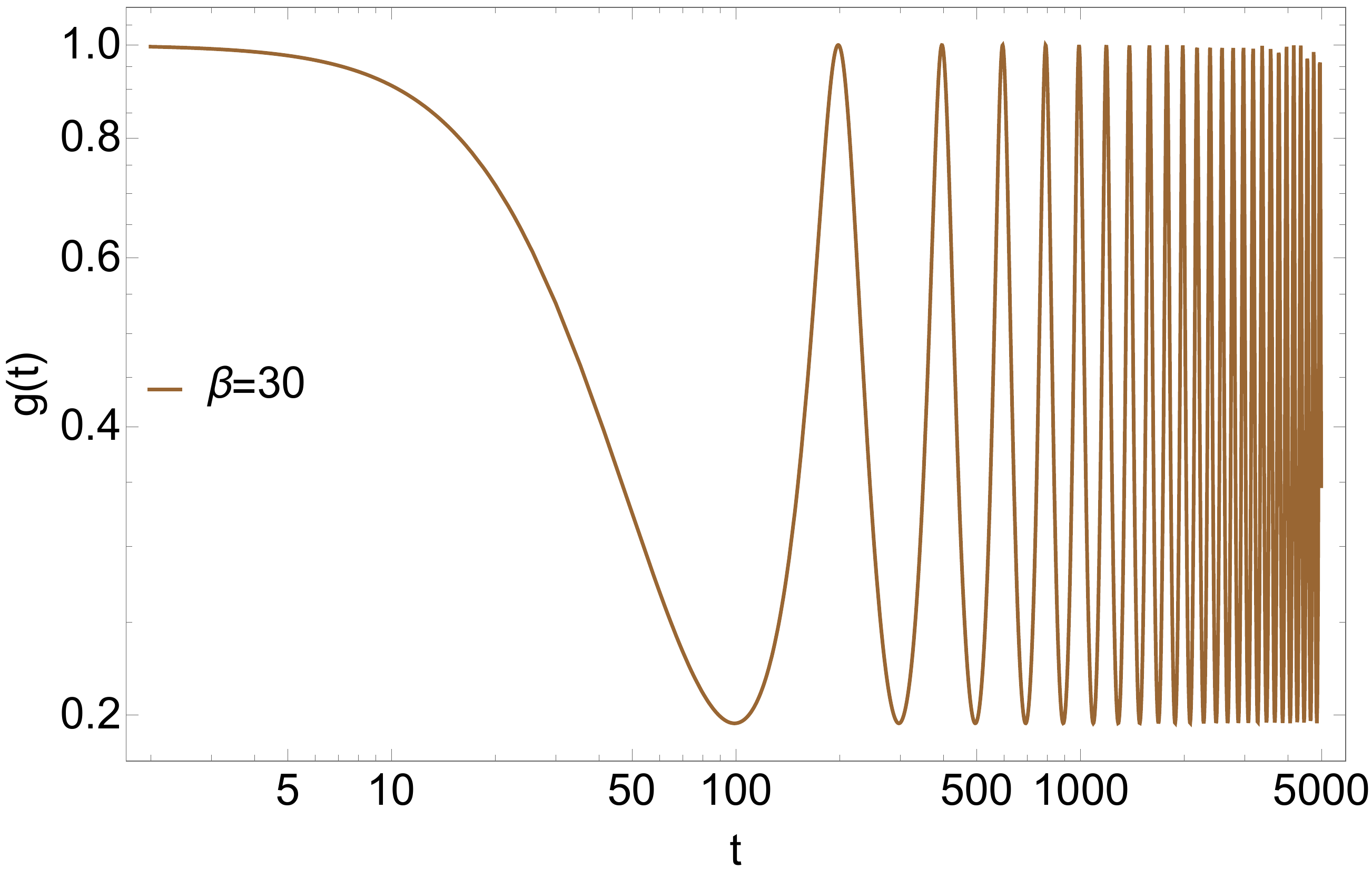}
    \end{subfigure}
    \hfill
    \begin{subfigure}{0.47\textwidth}
    \includegraphics[width=\textwidth]{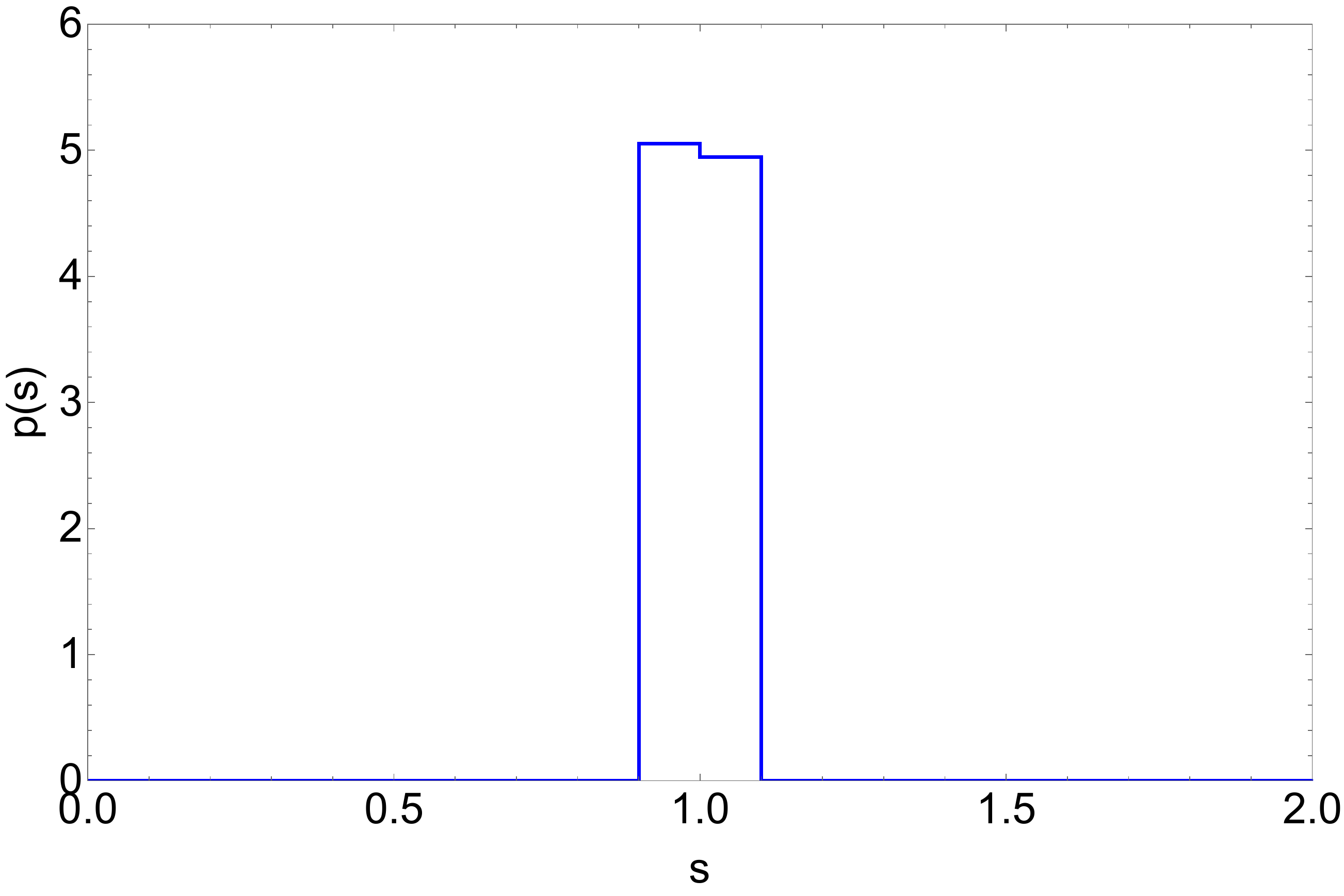}
    \end{subfigure}
    \caption{SFF (left) and corresponding level spacing distribution (right) for  BTZ black hole normal modes with $z_0=30$, $\beta=30$, $n_{cut}=200$ and $J$ is fixed to one. Bin size on the right is $0.1$.}
    \label{btz_n}
\end{figure}

\begin{figure}[H]
\begin{subfigure}{0.48\textwidth}
    \centering
    \includegraphics[width=\textwidth]{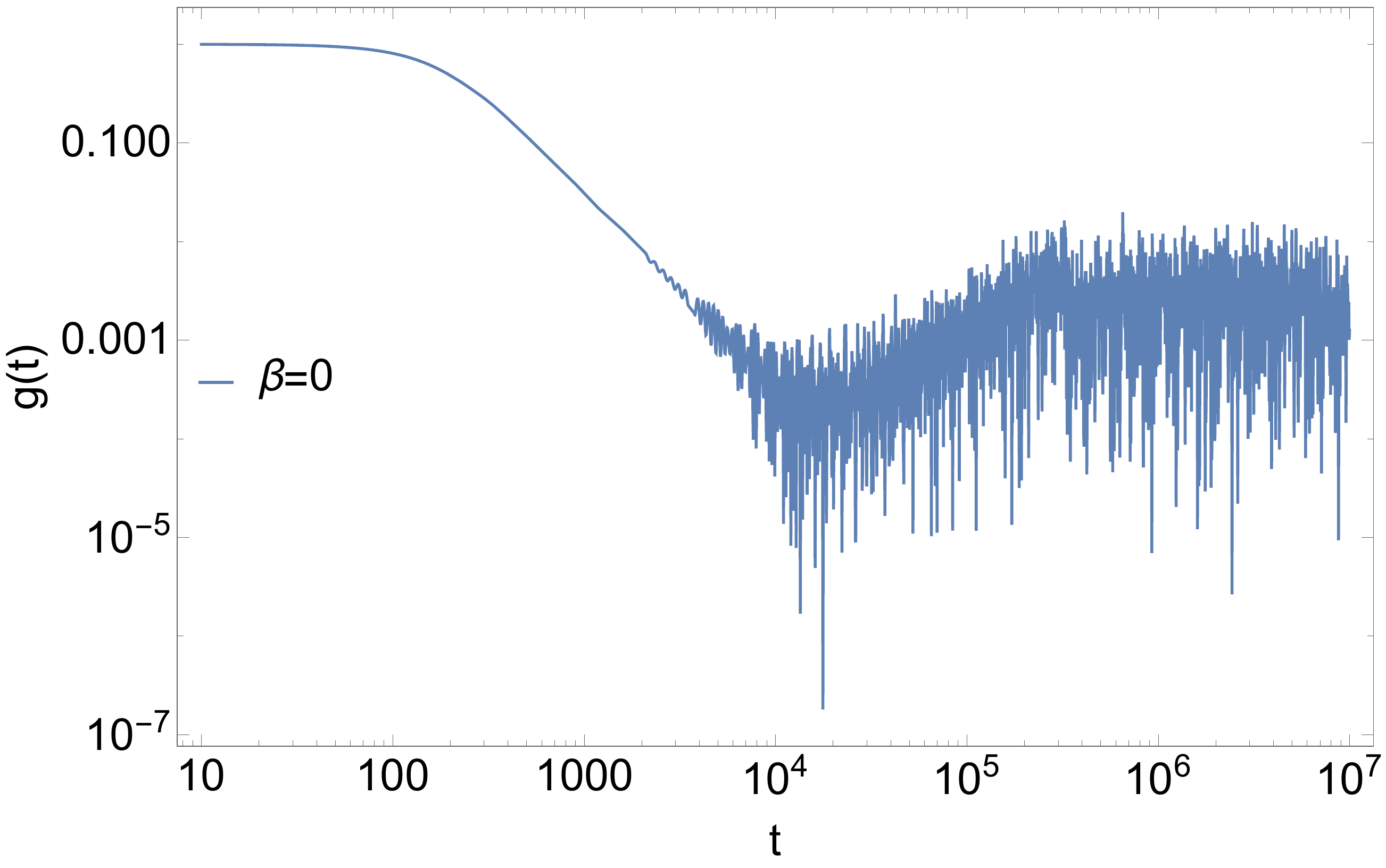}
    \end{subfigure}
    \hfill
    \begin{subfigure}{0.47\textwidth}
    \includegraphics[width=\textwidth]{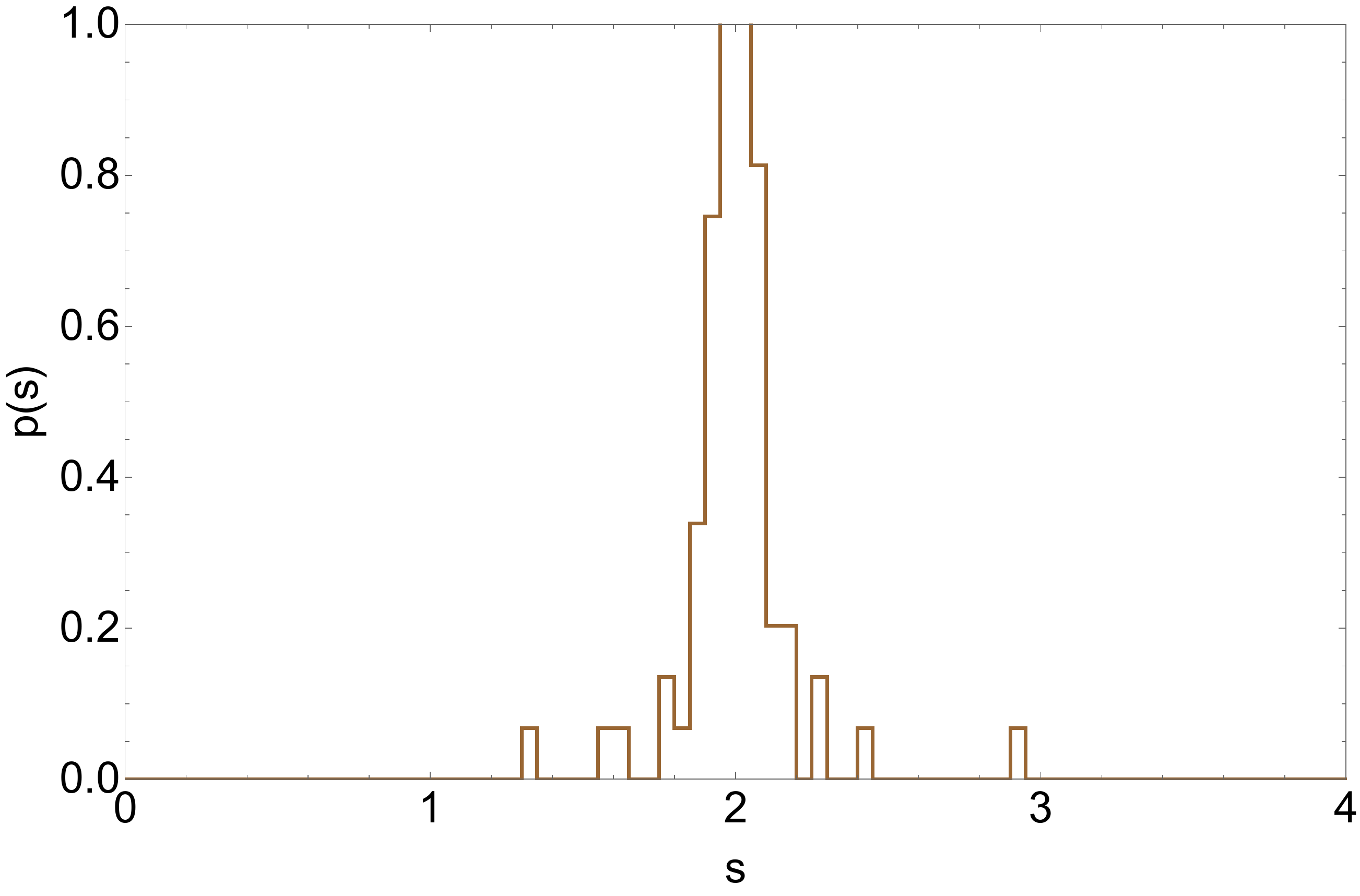}
    \end{subfigure}
    \caption{SFF (left) and corresponding level spacing distribution (right) for BTZ black hole normal modes with $z_0=30$, $\beta=0$, $J_{cut}=300$ and $n$ is fixed to one. Bin size on the right is $0.05$.}
    \label{btz_j}
\end{figure}

\begin{figure}[H]
\begin{subfigure}{0.48\textwidth}
    \centering
    \includegraphics[width=\textwidth]{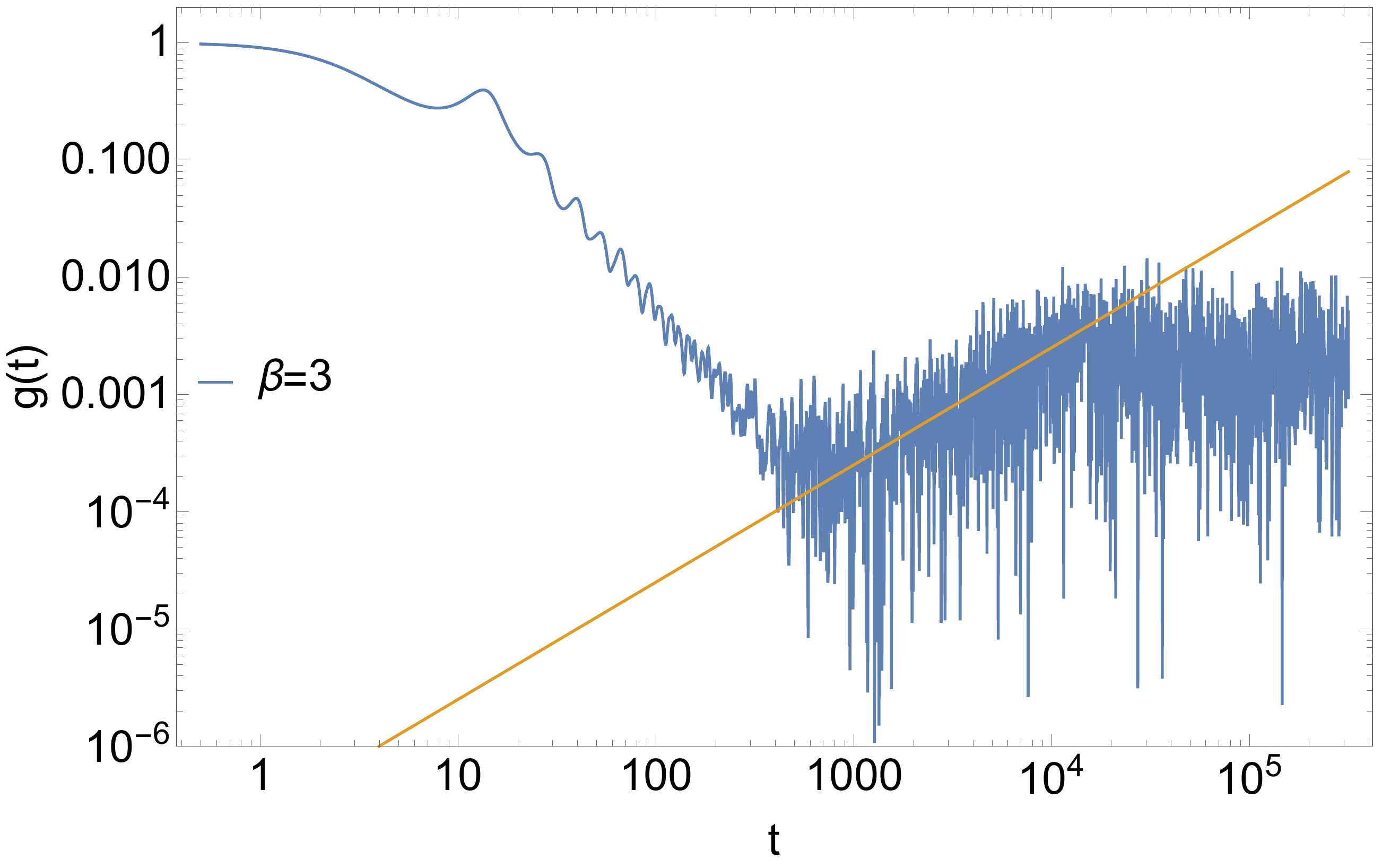}
    \end{subfigure}
    \hfill
    \begin{subfigure}{0.47\textwidth}
    \includegraphics[width=\textwidth]{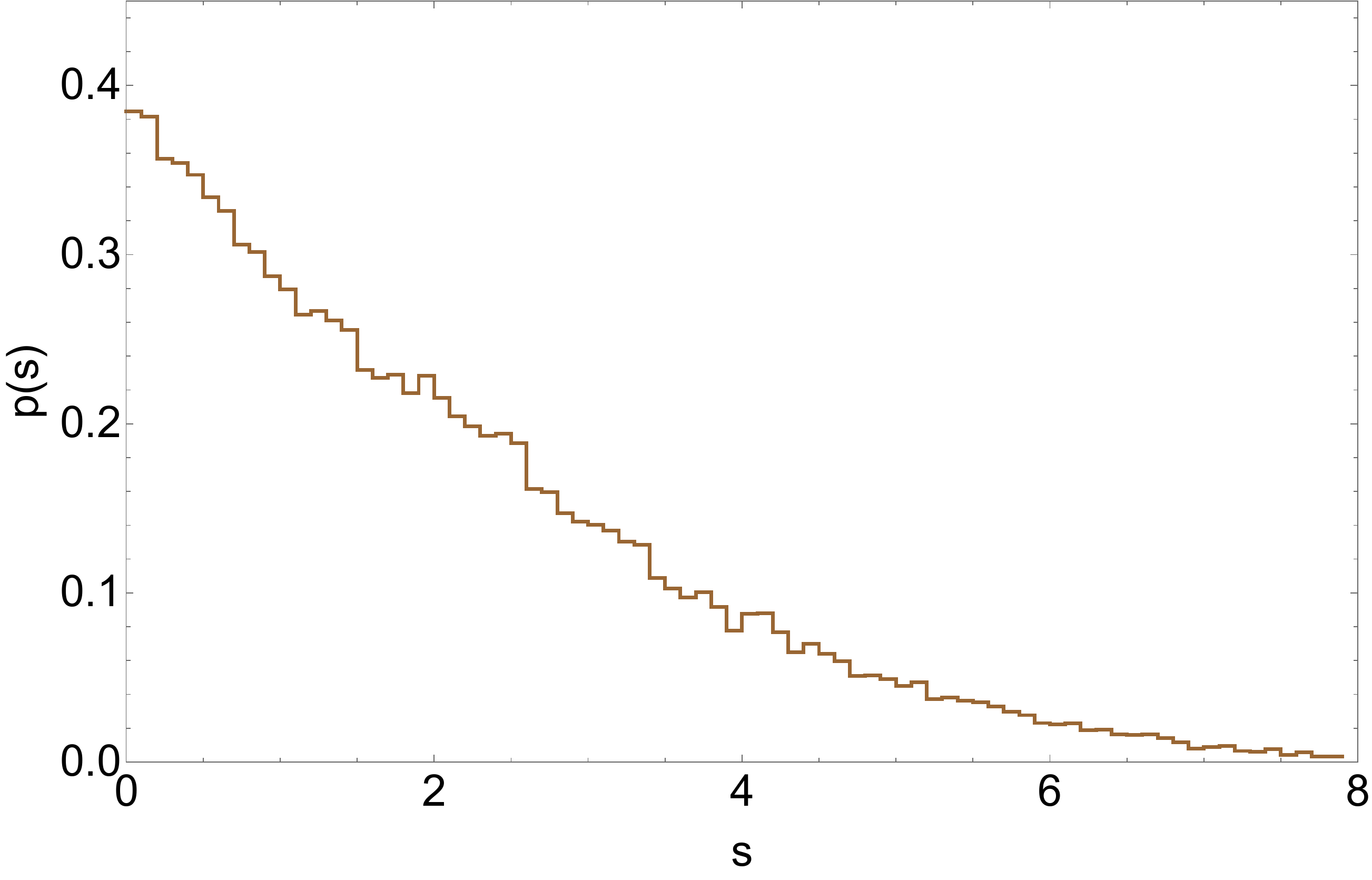}
    \end{subfigure}
    \caption{SFF (left) and its corresponding level spacing (right) for BTZ black hole normal modes with $z_0=12$, $n_{cut}=270$ and $J_{cut}=200$ at $\beta=3$. The yellow line has slope 1.}
    \label{btz_nj}
\end{figure}

In Figure \ref{btz_nj1} we have plotted the SFF $g(t)$ of the BTZ black hole for a choice of representative parameters: $\beta=30$, $J_{cut}=300$, and $n_{cut}=30$. The important point is that it shows Dip-Ramp-Plateau (DRP) structure without any averaging. We will see that this is {\em not} a generic feature of integrable systems.
 To see what is actually responsible for this DRP structure, we have separately plotted the SFF with only an $n$ sum (Figure \ref{btz_n}) and only a $J$ sum (Figure \ref{btz_j}).
 For a single $n$, it is evident that the SFF shows a clear DRP structure, but the ramp is completely lost if one includes only one $J$. So the $J$ sum is crucial. The right panel of Figure \ref{btz_j} shows the level spacing plot corresponding to fixed $n$ for a range of $J$ (with bin size $0.05$). It shows very clearly that its form is neither that of a conventional integrable system, nor of a standard RMT class. This is correlated with the SFF showing a DRP structure. If we include a sum over both $n$ and $J$ (ie., we work with large values of both $n_{cut}$ and $J_{cut}$) the level spacing plots look increasingly Poisson, but the DRP structure remains intact. This is illustrated by Figure \ref{btz_nj}. Clearly, the SFF is sensitive to the underlying level-spacing structure along the $J$-direction, even though the large $n_{cut}$ is drawing  out  the peaks in the level spacing. The physics of interest to us is visible already at $n_{cut}=1$ and this is the case that we will focus on in this paper.

\begin{figure}[H]
\begin{subfigure}{0.44\textwidth}
    \centering
    \includegraphics[width=\textwidth]{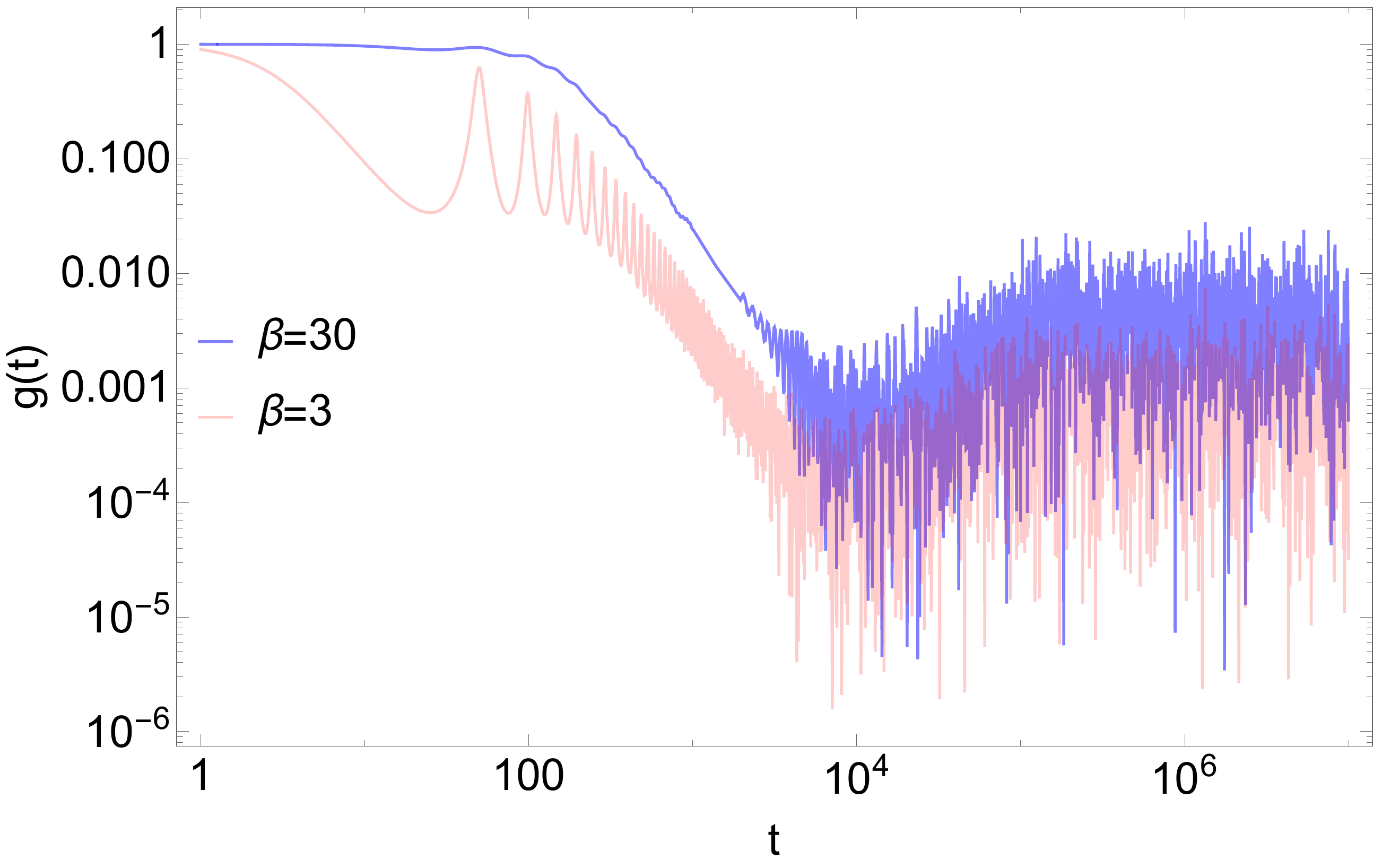}
    \end{subfigure}
    \hfill
    \begin{subfigure}{0.44\textwidth}
    \includegraphics[width=\textwidth]{fitting_ramp_btz_v1.pdf}
    \end{subfigure}
    \caption{Left: $\beta$ dependence of SFF  for BTZ black hole with $z_0=30$, $n_{cut}=200$ and $J_{cut}=200$. Right:  Ramp part of SFF for BTZ black hole ($z_0=30$, $\beta=0$, $J_{cut}=300$ and $n=1$) is contrasted against the yellow line whose equation is $\log{g(t)}=\log{t}+$constant.} 
    \label{btz_beta}
\end{figure}

 
The existence of the DRP structure is largely unaffected by the choice of $\beta$. There is a small caveat to this, namely that for small enough $\beta$ (ie., large enough temperature) and $n_{cut} > 1$,  the finiteness of the $n_{cut}$ in our calculations can give rise to some spurious features at the beginning of the dip. This is illustrated in Figure \ref{btz_beta} (left), and similar features can be produced in any system (say SHO) by introducing a cut off. This feature is naturally absent when $n_{cut}=1$, so we will usually focus on the $n_{cut}=1, \beta=0$ case. This would have merely been a technical observation about the numerics, except that this captures also a crucial fact -- that the dependence of energy eigenvalues on $J$ is much weaker than that in the $n$-direction. In other words, for a fixed value of $n$, the change in $\omega$ due to a change in $J$ is hierarchically smaller, than how it changes for a fixed $J$ with $n$. This quasi-degeneracy in the $J$-direction is directly correlated with the presence of the ramp. Note that this is  $not$ what happens for instance in global AdS where there is no horizon, and the dependence of energy eigenvalues on both $n$ and $J$ are of the same order.

\begin{figure}[H]
\begin{subfigure}{0.44\textwidth}
    \centering
    \includegraphics[width=\textwidth]{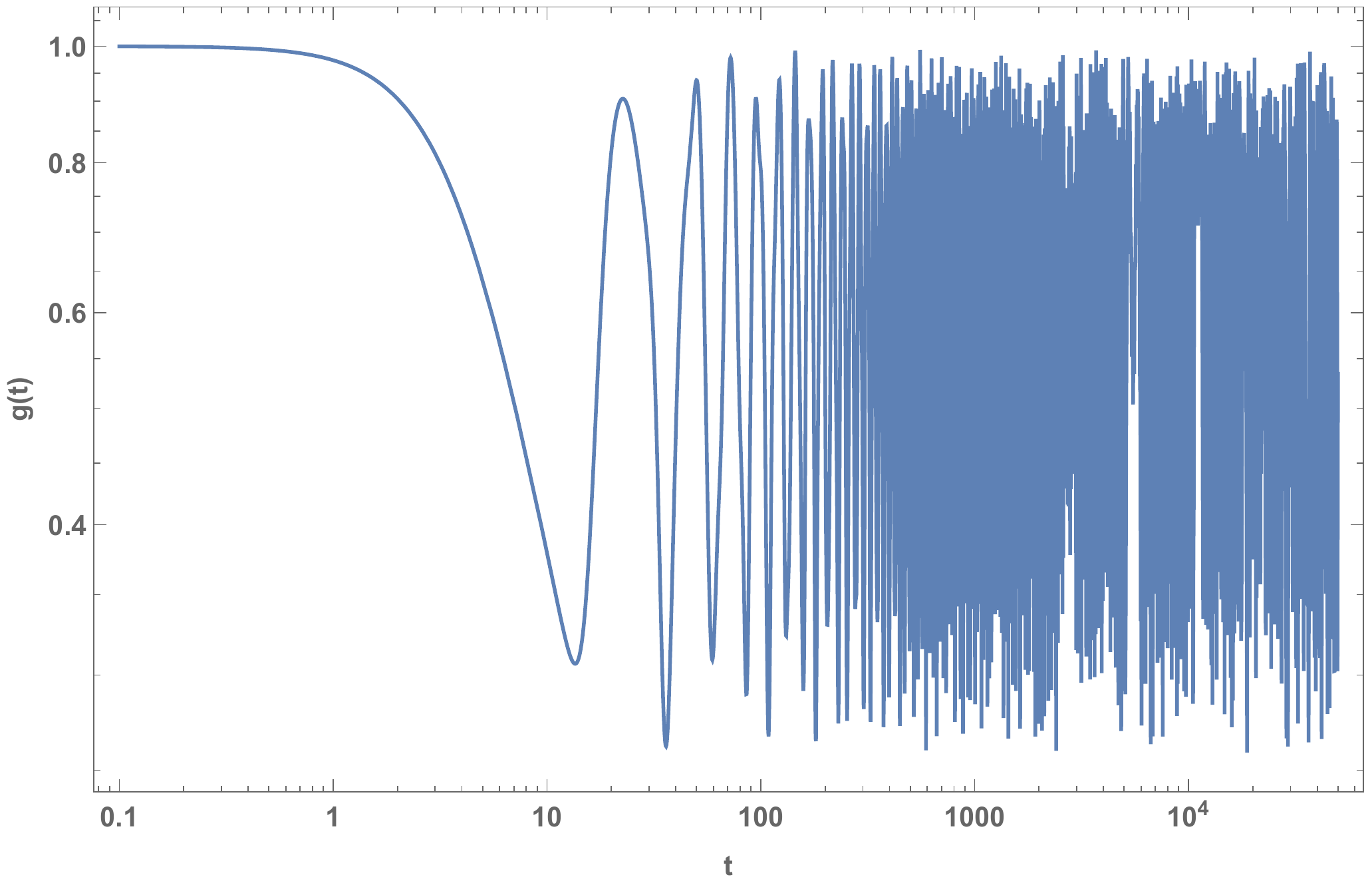}
    \end{subfigure}
    \hfill
    \begin{subfigure}{0.48\textwidth}
    \includegraphics[width=\textwidth]{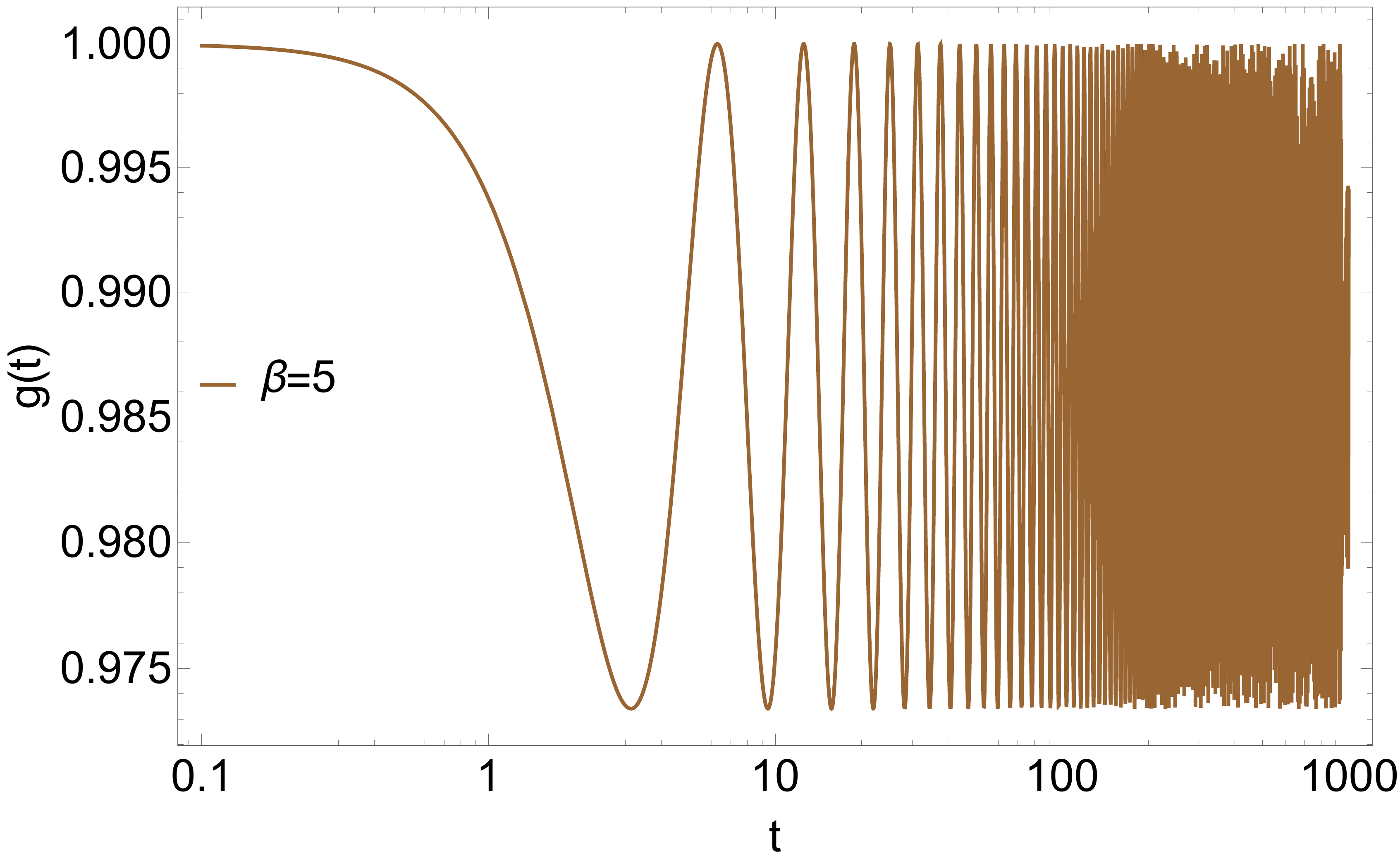}
    \end{subfigure}
    \caption{Left : SFF for BTZ black hole with $z_0=2$, $J_{cut}=80$, $n=1$ and $\beta=5$. Right: SFF for empty AdS with dispersion relation $\omega_{nl}=(\Delta+l+2n)$, where $\Delta=2$, $l_{cut}=100$, $n=1$ and $\beta=5$. These two figures show how the structure of SFF becomes empty AdS like when the cutoff is increased.}
    \label{btz_lowz}
\end{figure}

In Figure \ref{btz_beta} (right) we show that the slope of the ramp is consistent with unity, on a log-log plot. The slope of the ramp is stable as we change $J_{cut}$ and $n_{cut}$, even though we have not explored very large $n_{cut}$ values due to numerical difficulties. Our focus will be on features that are visible already at $n_{cut}=1$. In Figure \ref{btz_lowz} we show that the ramp structure vanishes when the stretched horizon is no longer close to the horizon. The SFF of global AdS is also presented in the same  figure for comparison; it also exhibits no ramp.  

\FloatBarrier
\begin{figure}[H]
    \centering
    \includegraphics[width=.46\textwidth]{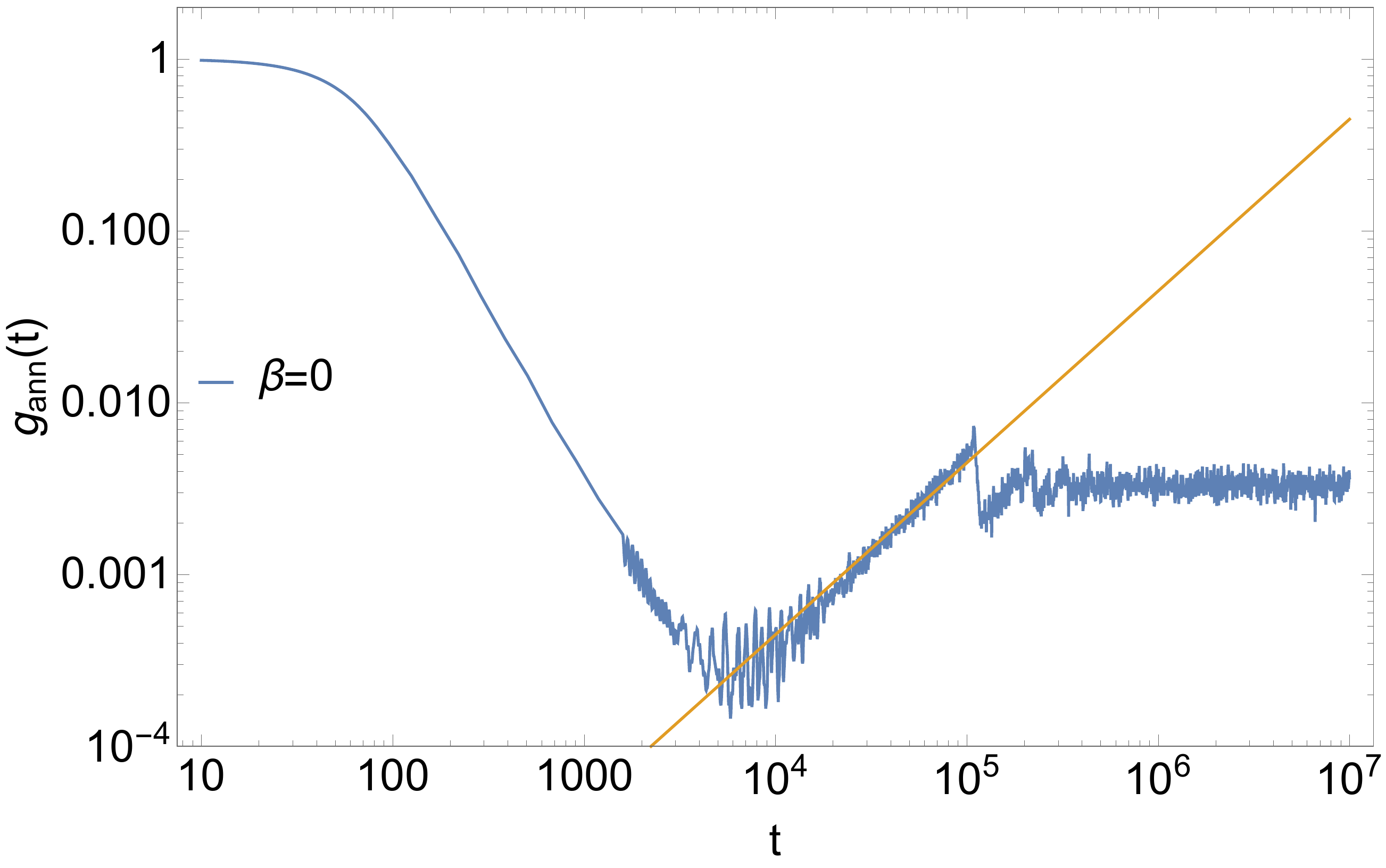}
   \caption{Annealed SFF of BTZ normal modes with $J_{cut}=300$, $\beta=0$ and $n=1$. Averaging is done over hundred randomly chosen $z_0$ from a normal distribution with mean $\mu=20$ and variance $\sigma=0.1$. The equation of yellow straight line is $\log g_{\text{ann}}(t)=\log t+$constant.} 
    \label{btz_many_realization}
\end{figure}
\FloatBarrier
%



We can also consider an averaged version of the SFF over an ensemble. A natural averaging in our case is over a collection of randomly varied $z_0$. We consider
\begin{equation}
    g_{\text{ann}}(t)=\frac{\langle {\rm Tr} (e^{-(\beta-i t)H}) {\rm Tr}(e^{-(\beta+i t)H}) \rangle
    }{\langle \left({\rm Tr}(e^{-\beta H})\right)^2\rangle
    }.
\end{equation}
For our BTZ problem, we take the average 
 over systems with different $z_0$ or equivalently $r_0$. For our plot in Figure \ref{btz_many_realization} we have taken hundred $z_0$ values from a normal distribution of mean $z_0=20$ and variance $\sigma=0.1$.
%
%


\section{Case Study II: Scalar Fields on the Rindler Wedge}

A ramp of slope unity is a surprising and apparently non-generic result in systems that do no exhibit level repulsion. 
So it is important to find out whether our results from the previous sections are a peculiarity specific to the BTZ geometry, or a generic feature of stretched horizons. In higher dimensions, solutions of black hole wave equations require Heun functions (which are not as well studied as the classical special functions). But a simple easily solvable candidate for stretched horizon can be obtained by considering the Rindler metric. This geometry has the extra virtue that the near horizon geometry of general non-extremal black holes is Rindler metric times some compact space. We will find that our observations of the last section remain intact in these systems as well, even though the details of the wave equation and the solutions\footnote{Solutions of Rindler wave equations are in terms of Bessel functions.} are quite different.  

\subsection{Warm up: $2$d Rindler}

We will start with 1+1 dimensions. Since the circle played a crucial role in the BTZ case in giving rise to the ramp, we expect that the 1+1 d Rindler will not exhibit a DRP structure. This will be our motivation for introducing compact dimensions into the game in the next subsection.

Starting with the two dimensional Minkowski metric, we can reach the Rindler wedge via the coordinate transformation
\begin{gather*}
     t =\frac{1}{a}e^{a\xi}\sinh{a\eta}, \hspace{0.5cm} x =\frac{1}{a}e^{a\xi}\cosh{a\eta},
\end{gather*}
where $a$ is a constant and $-\infty<\eta, \xi<\infty$. The result is
\begin{equation}
    ds^2=e^{2a\xi}(-d\eta^2+d\xi^2).
\end{equation}
which is conformal to Minkowski spacetime and has a horizon at $\xi\rightarrow-\infty$. Field equation for the massless scalar field $\Phi(\eta,\xi)$ becomes
\begin{equation}
   \left (-\partial_{\eta}^2+\partial_{\xi}^2\right)\Phi(\eta,\xi)=0.
\end{equation}
Solutions are of the form $\Phi=e^{-i \eta \omega} \phi(\xi)$ where $\phi(\xi)=C_1 \sin{\xi \omega}+C_2 \cos{\xi \omega}$. Now to compute normal modes we impose the boundary condition\footnote{In general, we will not care about the precise nature of the boundary conditions at the larger end of $\xi$. They do not qualitatively affect the discussion of the SFF. Here we have simply chosen to impose a Dirichlet condition at $\xi_0/2$, where $-\xi_0/2$ is the location of the stretched horizon. In higher dimensions, we will impose a vanishing condition at $\xi \rightarrow \infty$. The boundary condition at the stretched horizon on the other hand, will be chosen as Dirichlet in all dimensions.} $\phi\left(\pm\xi_0/2\right)=0$, which implies that $\omega$'s are quantized and given by $\omega_n=\frac{2n \pi}{\xi_0}$ where $n=1,2,3, \ldots$.

\begin{figure}[h]
    \centering
    \includegraphics[width=.55\textwidth]{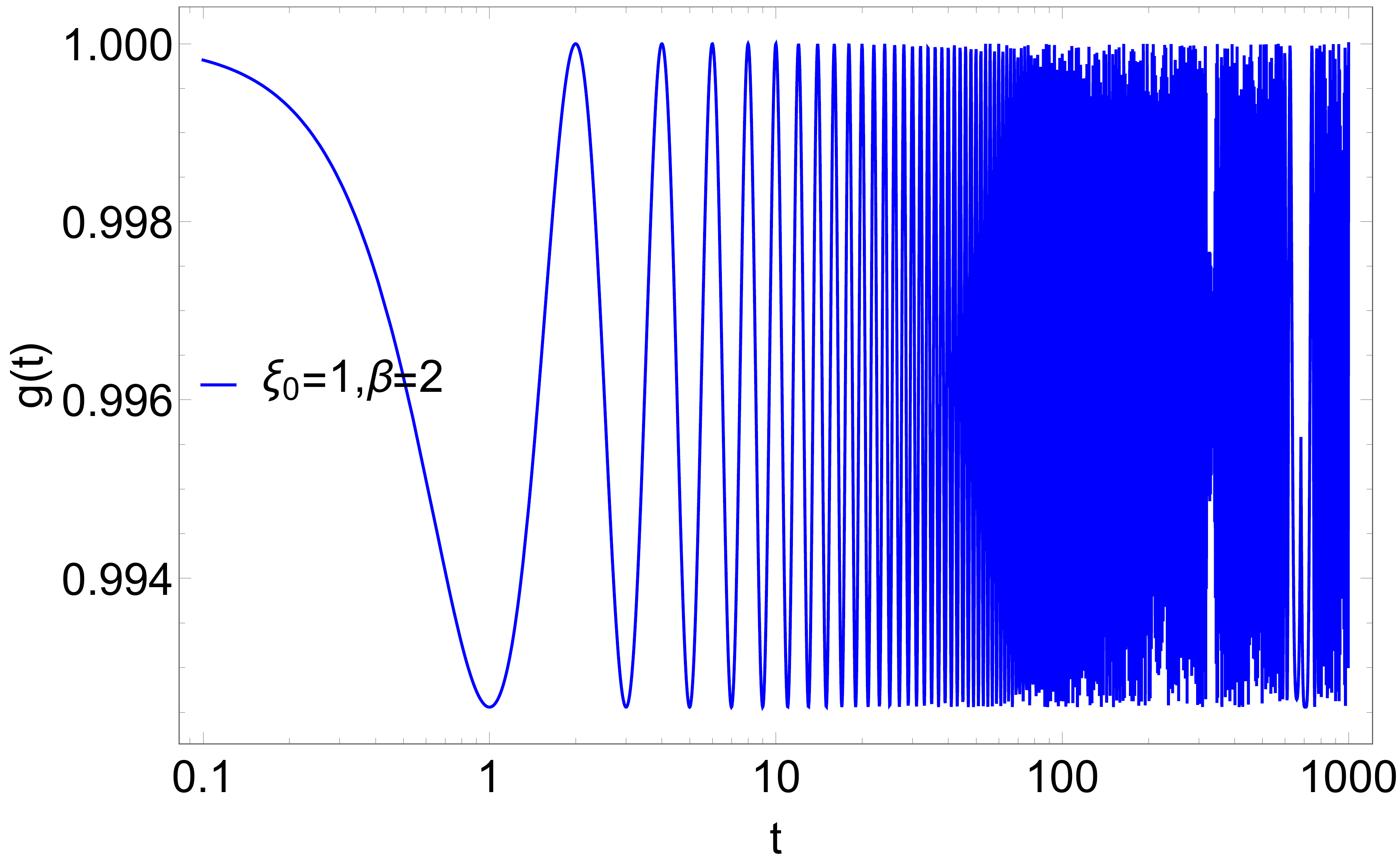}
    \caption{SFF for $2d$ Rindler normal modes with $\xi_0=1$ and $\beta=2$.}
    \label{sff_rindler1}
\end{figure}

With these normal modes we can compute the spectral form factor. It takes the form
\begin{align}\label{sffrindler}
    g(\beta,t)    
    &=\frac{1}{1+\sin^2{\frac{\pi t}{\xi_0}}\text{csch}^2{\frac{\pi \beta}{\xi_0}}}.
\end{align}
Figure \ref{sff_rindler1} shows the behaviour of the SFF, clearly there is no DRP structure. 

%
%


\subsection{2d Rindler $\times S^1$}

To see an interesting DRP structure, in the BTZ case we needed the dependence of the modes on the quantum number $J$ associated to the circle. This suggests that compact directions are essential for the emergence of the ramp. Indeed, we will now show that introducing a compact circle into the Rindler metric\footnote{Replacing the circle with another compact space does not affect the following discussion. The only difference is that the the quantum number $J$ is replaced by a suitably defined Casimir of the compact geometry.} 
\begin{equation}
    ds^2=e^{2 a \xi}(-d\eta^2+d\xi^2)+R^2d\psi^2
\end{equation}
results in an SFF with a DRP. Plugging in the form $\Phi(\xi,\eta,\psi)=e^{-i\eta\omega} e^{i\psi J} \phi(\xi)$ into the scalar field equation leads to 
\begin{equation}\label{rindler3eom}
    -e^{2 a \xi} J^2 \phi(\xi)+R^2\left(\omega^2 \phi(\xi)+\phi''(\xi)\right)=0.
\end{equation}
The general solution is given in terms of modified Bessel functions $I_{\alpha}(x)$ (which are written below as $I[\alpha,x]$):
\begin{equation}\label{rindler3d}
    \phi(\xi)=(-1)^{-\frac{i\omega}{2a}}\Tilde{C_1} \Gamma(1-\frac{i\omega}{a}) I\Big[-\frac{i\omega}{a},e^{a\xi}\frac{J}{a R}\Big]+(-1)^{\frac{i\omega}{2a}}\Tilde{C_2} \Gamma(1+\frac{i\omega}{a}) I\Big[\frac{i\omega}{a},e^{a\xi}\frac{J}{a R}\Big].
\end{equation}
Here $\Tilde{C_1}$ and $\Tilde{C_2}$ are arbitrary constants. $J=0$ limit of the above solution \eqref{rindler3d} is not well defined but from \eqref{rindler3eom} it is clear that solution for $J=0$ is in terms of usual oscillatory functions. We will see that demanding dying fields in the asymptotic region forces us to exclude the $J=0$ mode. 



For convenience, we rewrite equation \eqref{rindler3d} as
\begin{equation}\label{rind3}
    \phi(y)=C_1 I[-i A,y]+C_2 I[i A,y],
\end{equation}
with $A\equiv \omega/a$ and $y\equiv e^{a \xi}(J/a R)$. In terms of $y$ variable the position of boundary and horizon are given by $y\rightarrow\infty$ and $y\rightarrow0$ respectively. We will demand the boundary condition that the field $\phi(y)$ vanishes at boundary and is zero at some small $y_0$ which corresponds to our  stretched horizon, which we take to be at $\xi=-\xi_0$.
Near $y\rightarrow\infty$, equation \eqref{rind3} can be approximated as,
\begin{equation}
    \phi(y)\rightarrow(C_1+C_2)\frac{e^y}{\sqrt{2\pi y}}+(C_1 e^{\pi A}+C_2 e^{-\pi A})\frac{e^{-y}}{\sqrt{2\pi y}}.
\end{equation}
So vanishing condition at the boundary implies $C_1=-C_2$. Then boundary condition at $y_0$ implies,
\begin{align}
    &\phi(y_0)=C_1(I[-i A,y_0]-I[i A,y_0])=0\nonumber \\
     &\implies I[-i A,y_0]=I[i A,y_0]
\end{align}
Near horizon i.e. in the limit $y_0\rightarrow0$ the above expressions can be approximated as
\begin{align*}
    y_0^{-i A}\frac{2^{i A}}{\Gamma[1-i A]} &=y_0^{i A}\frac{2^{-i A}}{\Gamma[1+i A]}\\
    \implies\frac{\Gamma[i A]}{\Gamma[-i A]} &=-y^{2 i A} 2^{-2i A}\\
    &=-\Big(\text{Sgn}(y_0) |y_0|\Big)^{2 i A} 2^{-2 i A}\\
    &=-\big(\text{Sgn}(J)\big)^{2 i A} |y_0|^{2 i A} 2^{-2 i A}\\
    &=-\big(\text{Sgn}(J)\big)^{2 i A} e^{2 i A \log(|y_0|/2)}
\end{align*}
Equating the arguments of the both sides and substituting back the expressions for  $A_0$ and $y_0$ we finally have,
\begin{equation}\label{quantrind3}
     \frac{2\omega}{a} \log\frac{e^{-a \xi_0}J}{2 a R}+\text{Arg}\Big[(\text{Sgn}(J))^{ \frac{2 i\omega}{a}}\Big]+\pi=2\text{Arg}\Big[\Gamma\Big(\frac{i \omega}{a}\Big)\Big].
\end{equation}
The above equation \eqref{quantrind3} determines the allowed normal mode frequencies for this Rindler geometry. Given $R, \xi_0$ and $a$, the normal modes $\omega(n, J)$ can be numerically solved for as a function of $J$ (and an integer $n$). One point worth repeating is that $J=0$ modes are not allowed because the corresponding solution is incompatible with the boundary conditions. This can be viewed as a remnant of the fact that the $J=0$ mode corresponds to a dimensionally reduced problem.

\begin{figure}[H]
    \centering
    \includegraphics[width=.6\textwidth]{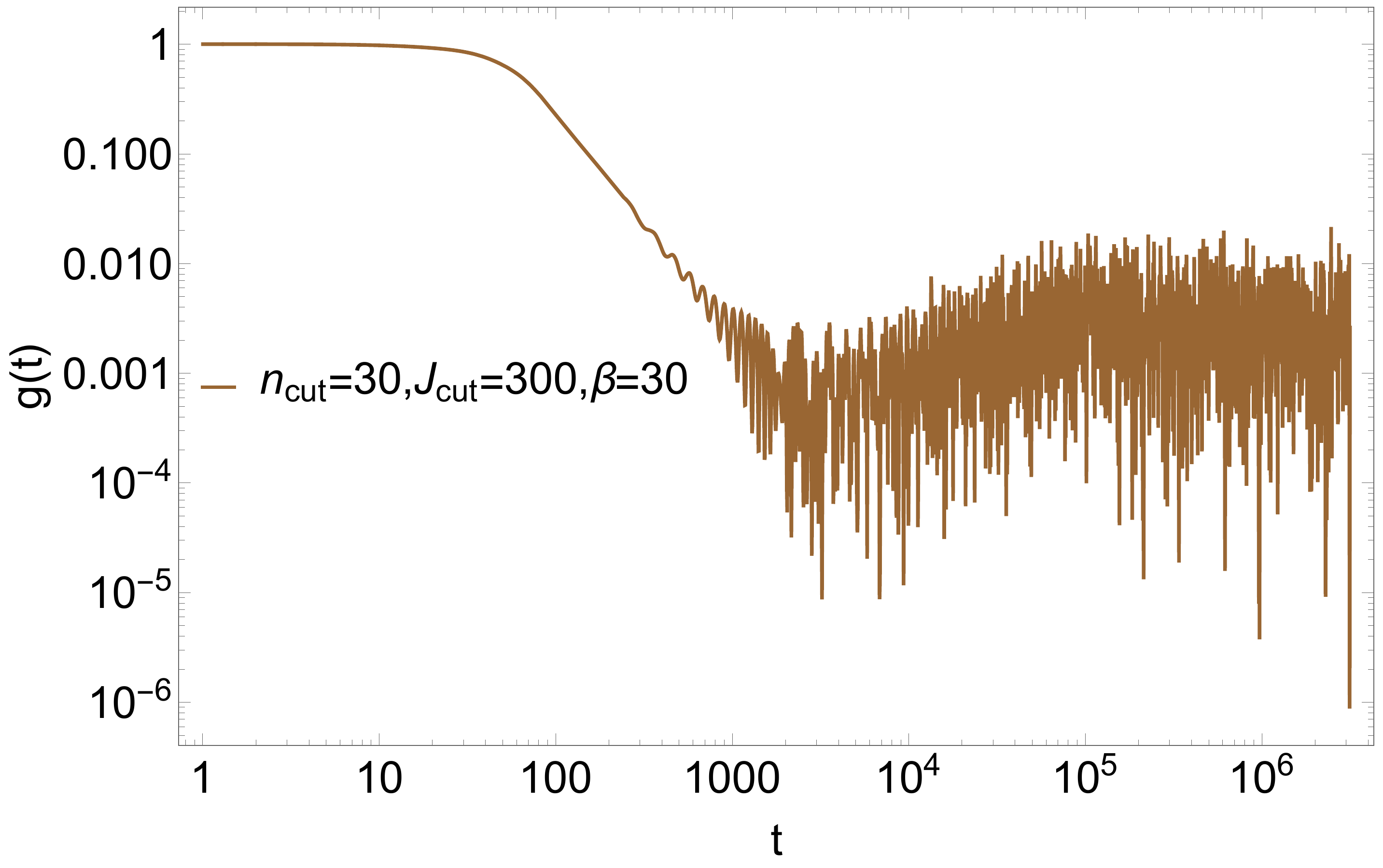}
    \caption{SFF for $3d$ Rindler normal modes with $n_{cut}=30$, $J_{cut}=300$, $\beta=30$. Other parameters are fixed to $\xi_0=20$, $R=2$, $a=1$.}
    \label{nj_rindler3}
\end{figure}

Figure\ref{nj_rindler3} shows a typical plot of the SFF for this Rindler geometry. The qualitative features we find in the Rindler case are {\em identical} to what we reported in the previous section, so we will not repeat the discussion here -- we will simply refer to the plots. The slogan we take away is that the general features that we saw for the SFF in the BTZ case are likely universal features of stretched horizons. 

\FloatBarrier
%


%
\begin{figure}[H]
\begin{subfigure}{0.46\textwidth}
    \centering
    \includegraphics[width=\textwidth]{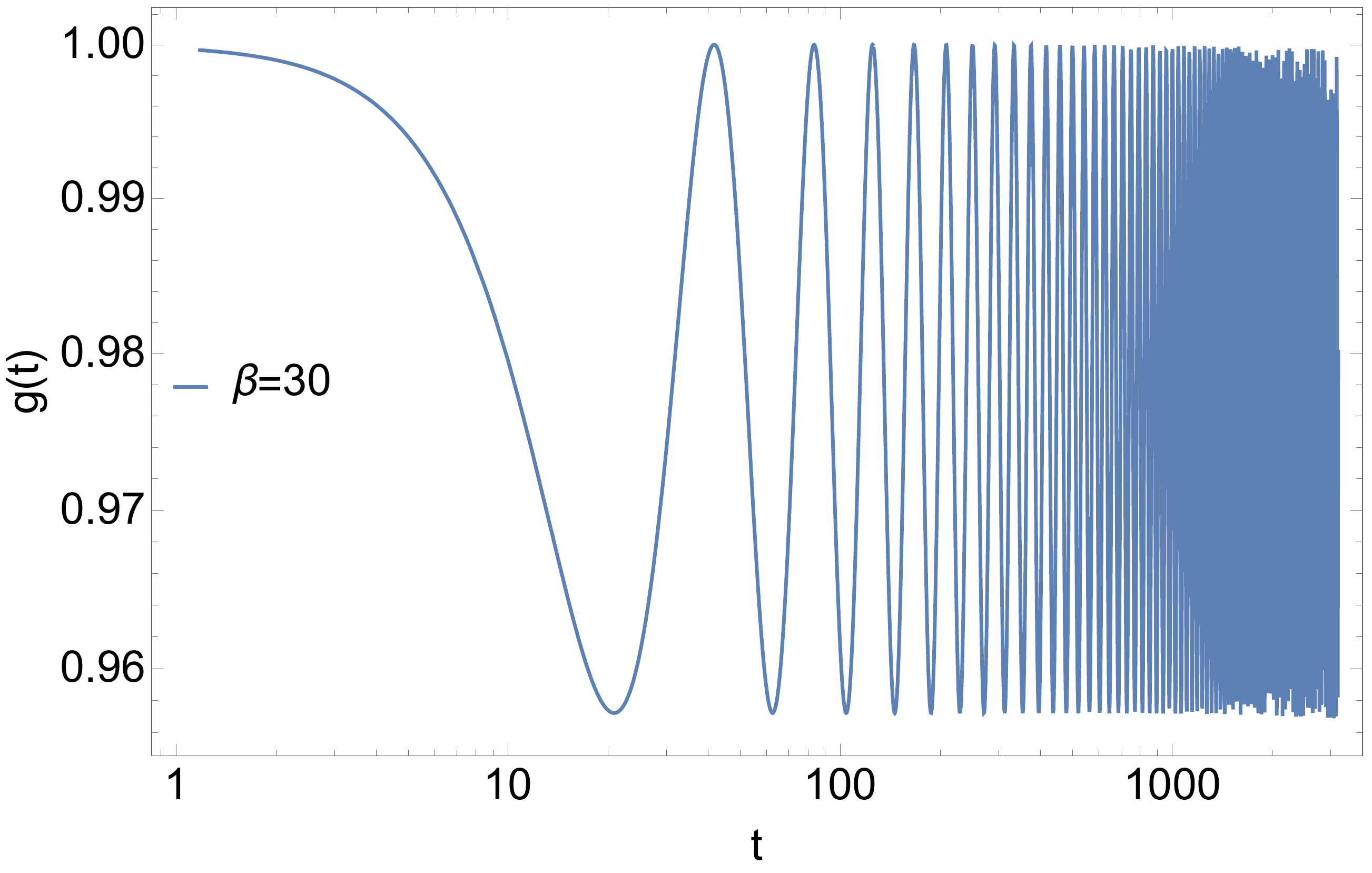}
    \end{subfigure}
    \hfill
    \begin{subfigure}{0.46\textwidth}
    \includegraphics[width=\textwidth]{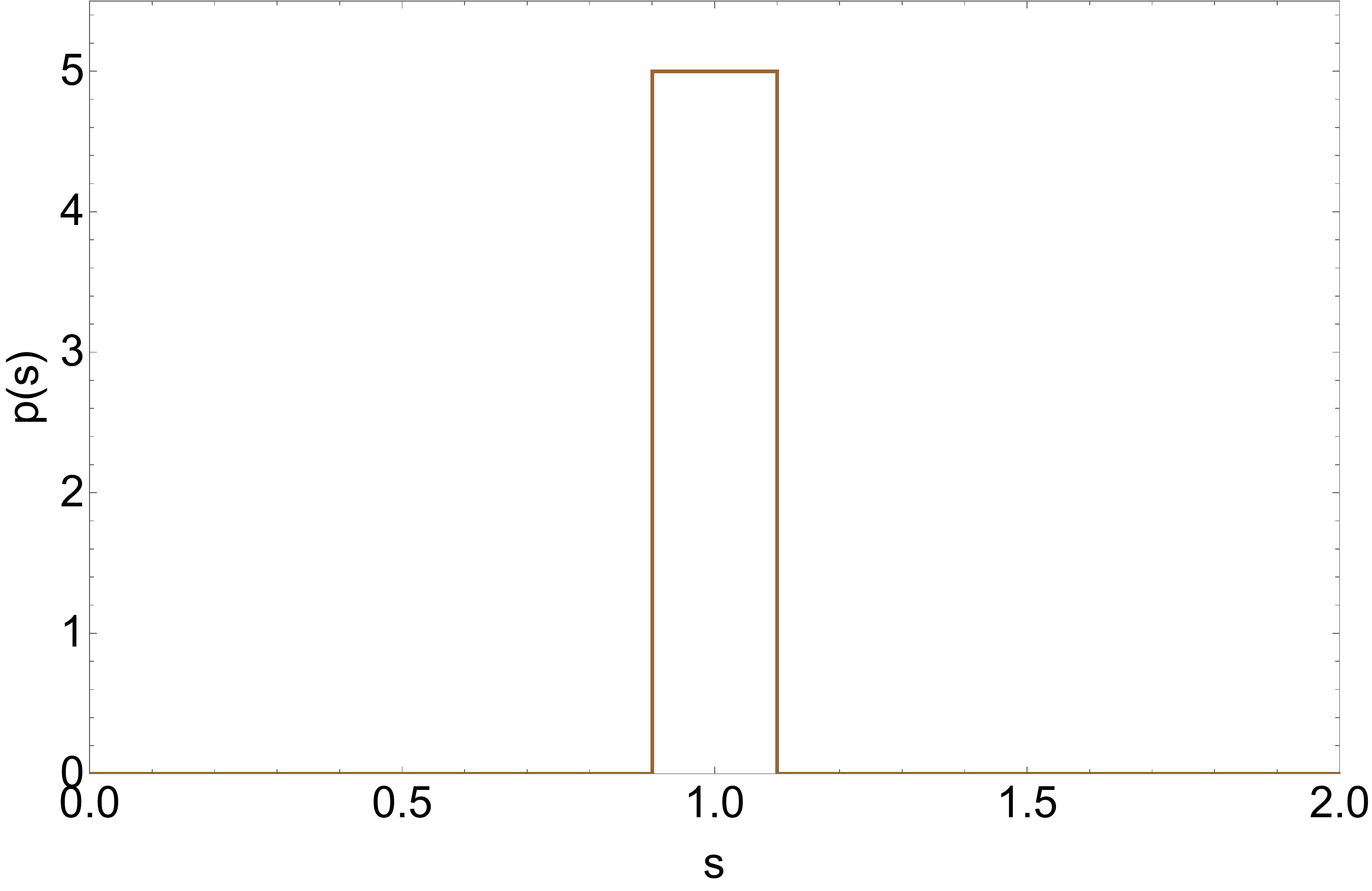}
    \end{subfigure}
    \caption{SFF (left) and level spacing distribution (right) for $3d$ Rindler normal modes with only $n$ sum (with $\xi_0=20$, $n_{cut}=200$ and $J$ is fixed to one) at $\beta=30$. Other parameters $R$ and $a$ are fixed to  $2$ and $1$. The bin size on the right is 0.1.}
    \label{sff_rindler3_n}
\end{figure}
\begin{figure}[H]
\begin{subfigure}{0.46\textwidth}
    \centering
    \includegraphics[width=\textwidth]{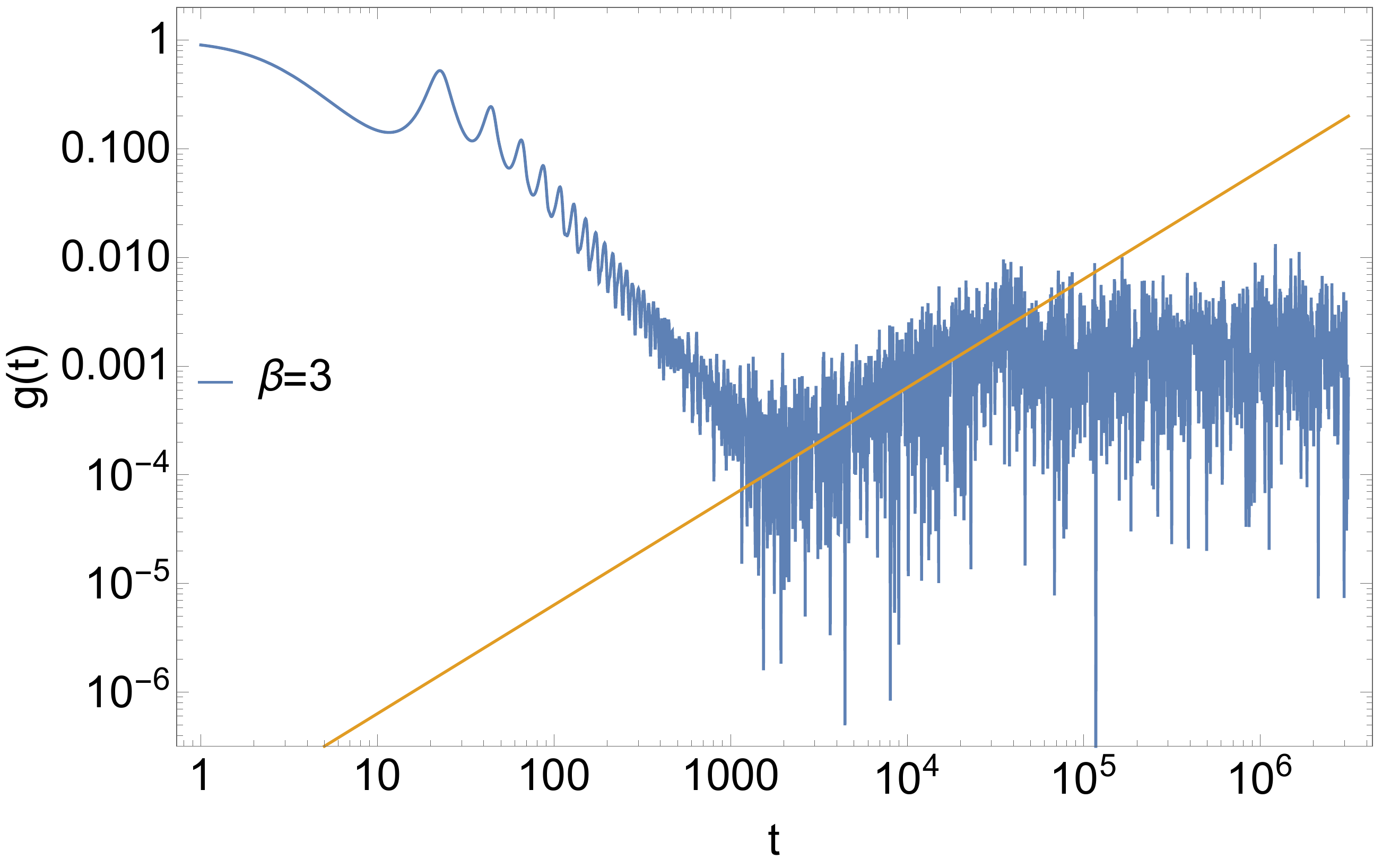}
    \end{subfigure}
    \hfill
    \begin{subfigure}{0.46\textwidth}
    \includegraphics[width=\textwidth]{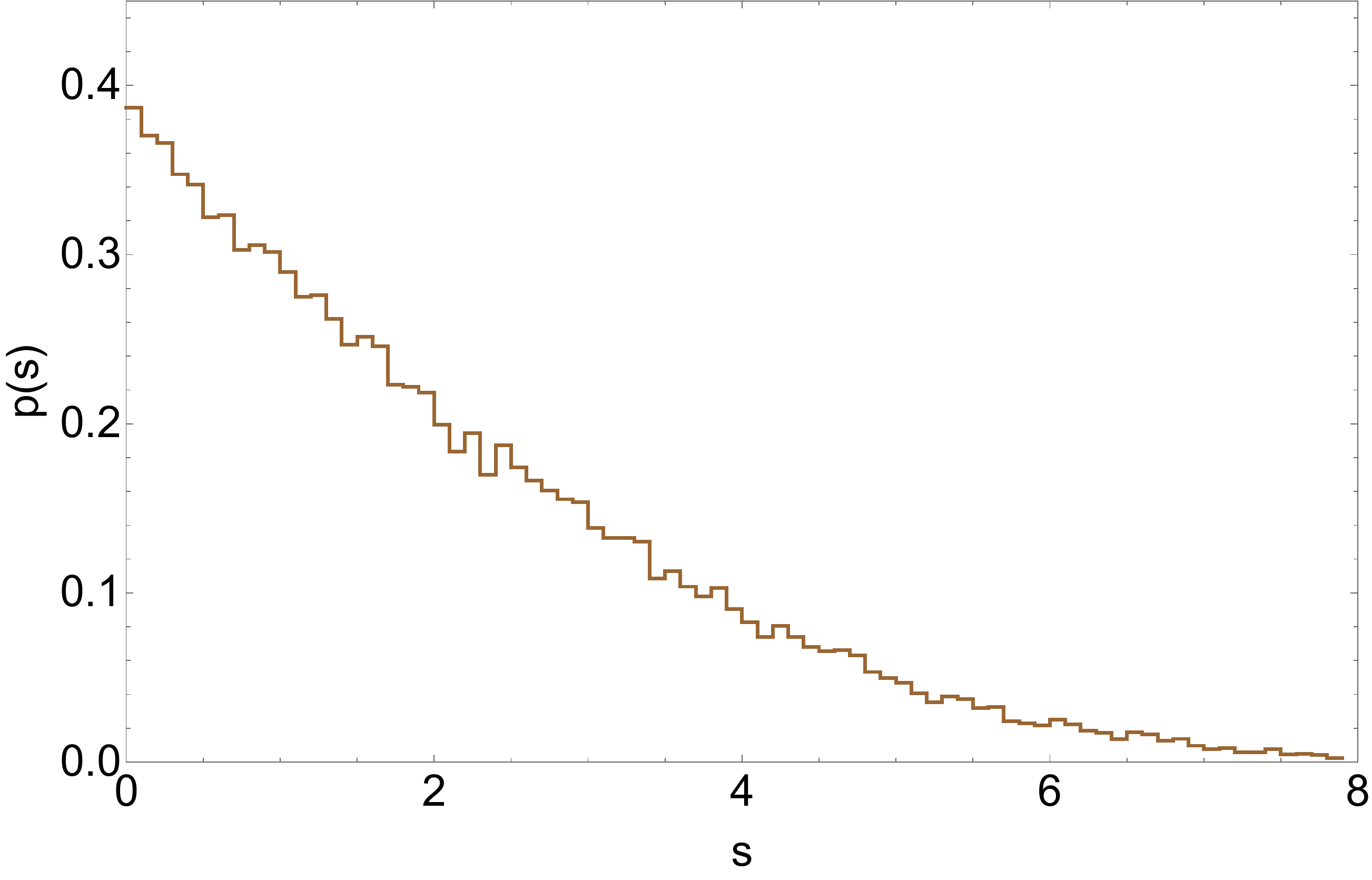}
    \end{subfigure}
    \caption{SFF (left) and its corresponding level spacing (right) for $3d$ Rindler normal modes with $\xi_0=15$, $n_{cut}=300$, $J_{cut}=200$, $\beta=3$ and $R=2$, $a=1$. Equation of the yellow straight line on the left is $\log g(t)=\log t+\text{constant}$.}
    \label{rind3_nj}
\end{figure}
\begin{figure}[H]
\begin{subfigure}{0.48\textwidth}
    \centering
    \includegraphics[width=\textwidth]{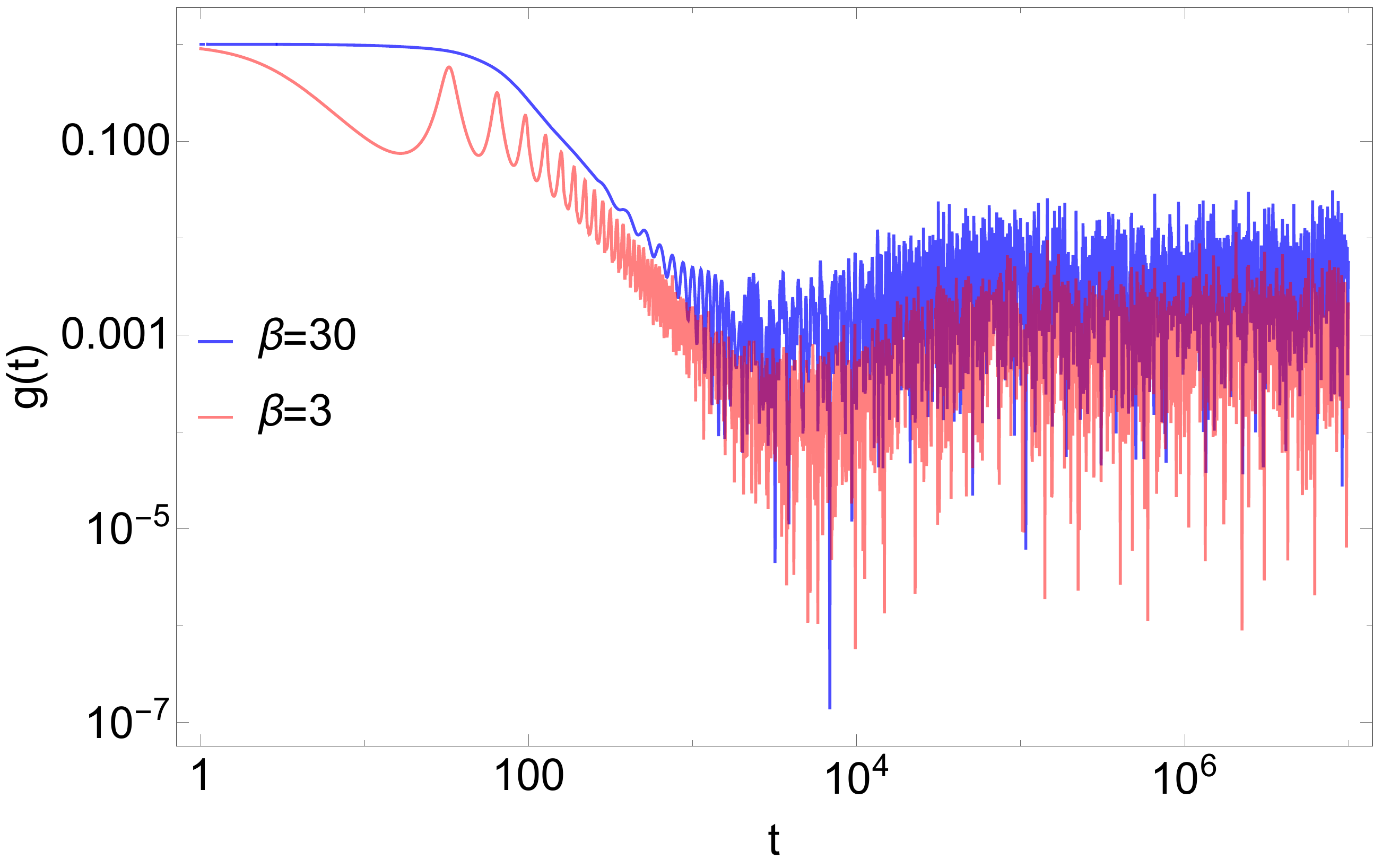}
    \end{subfigure}
    \hfill
    \begin{subfigure}{0.47\textwidth}
    \includegraphics[width=\textwidth]{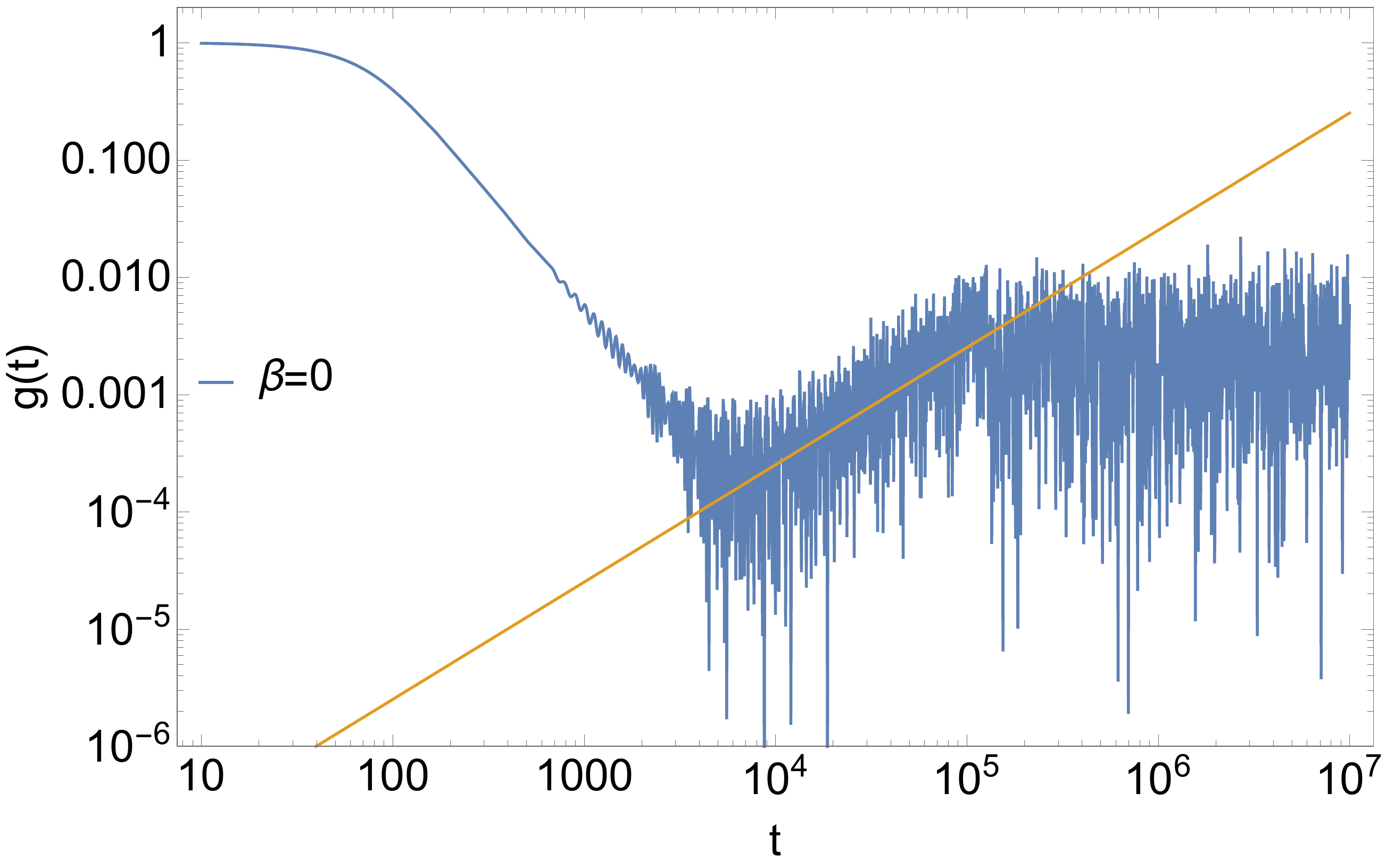}
    \end{subfigure}
    \caption{Left: $\beta$ dependence of SFF  for $3$d Rindler normal modes with $\xi_0=20$, $R=2$, $a=1$, $n_{cut}=200$ and $J_{cut}=200$. Right: Ramp part of SFF for $3$d Rindler normal modes with only $J$ sum ($n=1$, $\xi_0=20$, $\beta=0$ and $J_{cut}=300$, $R=2$, $a=1$) is contrasted against a yellow straight line whose equation is $\log{g(t)}=\log{t}+$constant.
    }
    \label{sff_rind3_temp}
\end{figure}
%
%
%
%
%
\begin{figure}[H]
    \centering
    \includegraphics[width=.5\textwidth]{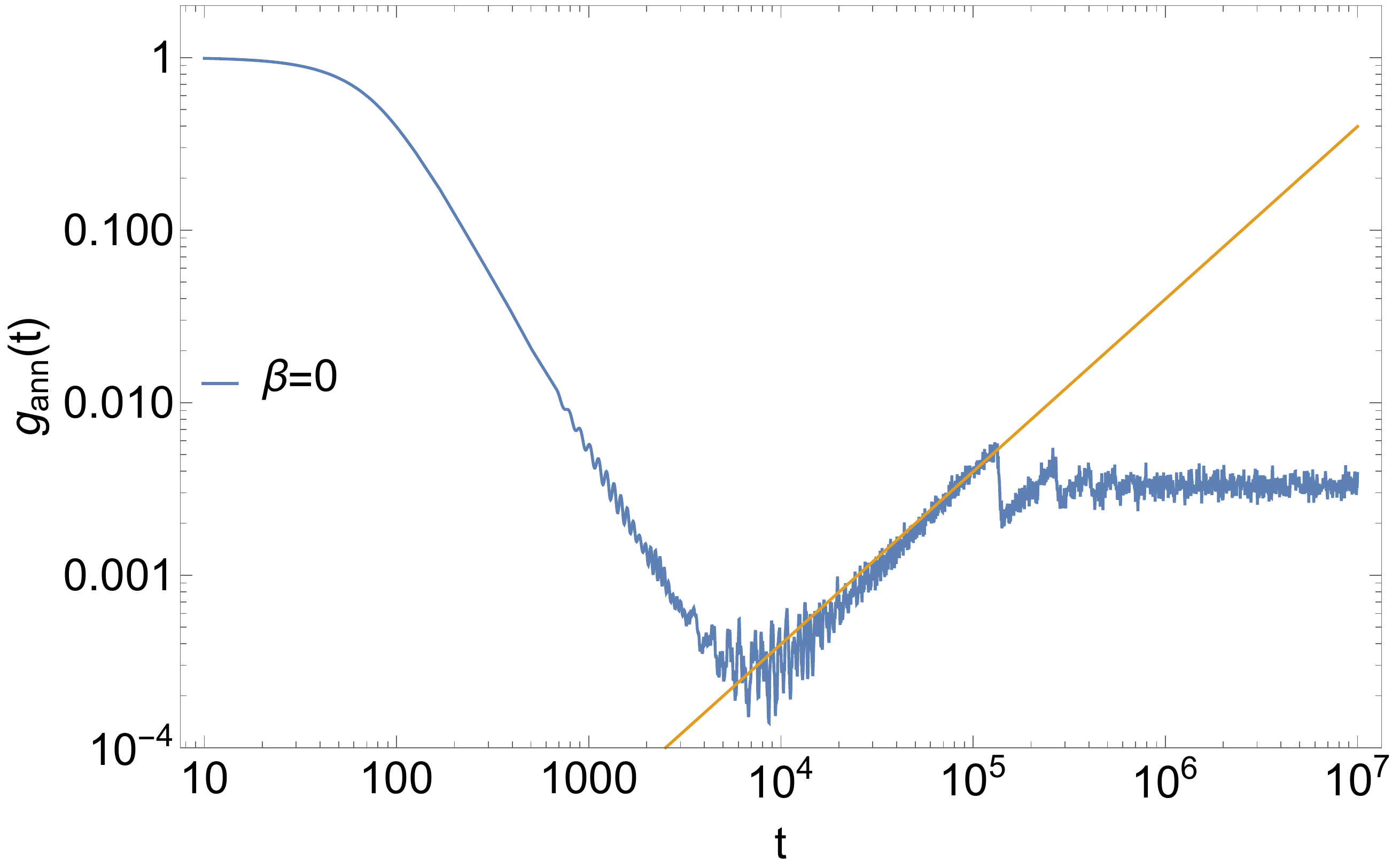}
    \caption{Annealed SFF for $3$d Rindler normal modes with fixed $n=1$, $\beta=0$, $a=1$, $R=2$ and $J_{cut}=300$. Averaging is done over hundred randomly chosen $\xi_0$ from a normal distribution with mean $\mu=20$ and variance $\sigma=0.1$. The equation of yellow straight line in the figure is $\log g_{\text{ann}}(t)=\log t$+constant.}
    \label{sff_rind3_avg}
\end{figure}
\FloatBarrier
%

\section{(No) Ramp in Typical Integrable Models}

In the previous section, we saw that the DRP structure of the SFF is likely a general feature of stretched horizons. In this section, we will show that despite this, this is {\em not} a generic feature of integrable systems. The examples we will consider are\footnote{We will emphasize integrable systems with Poisson level spacing. Of course, examples like SHO are also integrable systems that exhibit no ramp, and therefore consistent with our observation.} -- integrable rectangular billiards, SYK$_2$ model with random couplings, and an integrable system that we call the square-mod system. 
All of these cases are integrable, and in all of them the level spacing is known to be Poisson. We will present the level spacing plots along with their SFF to demonstrate the absence of ramp.




\subsection{Rectangular Billiard}

\begin{figure}[h]
\begin{subfigure}{0.5\textwidth}
    \centering
    \includegraphics[width=\textwidth]{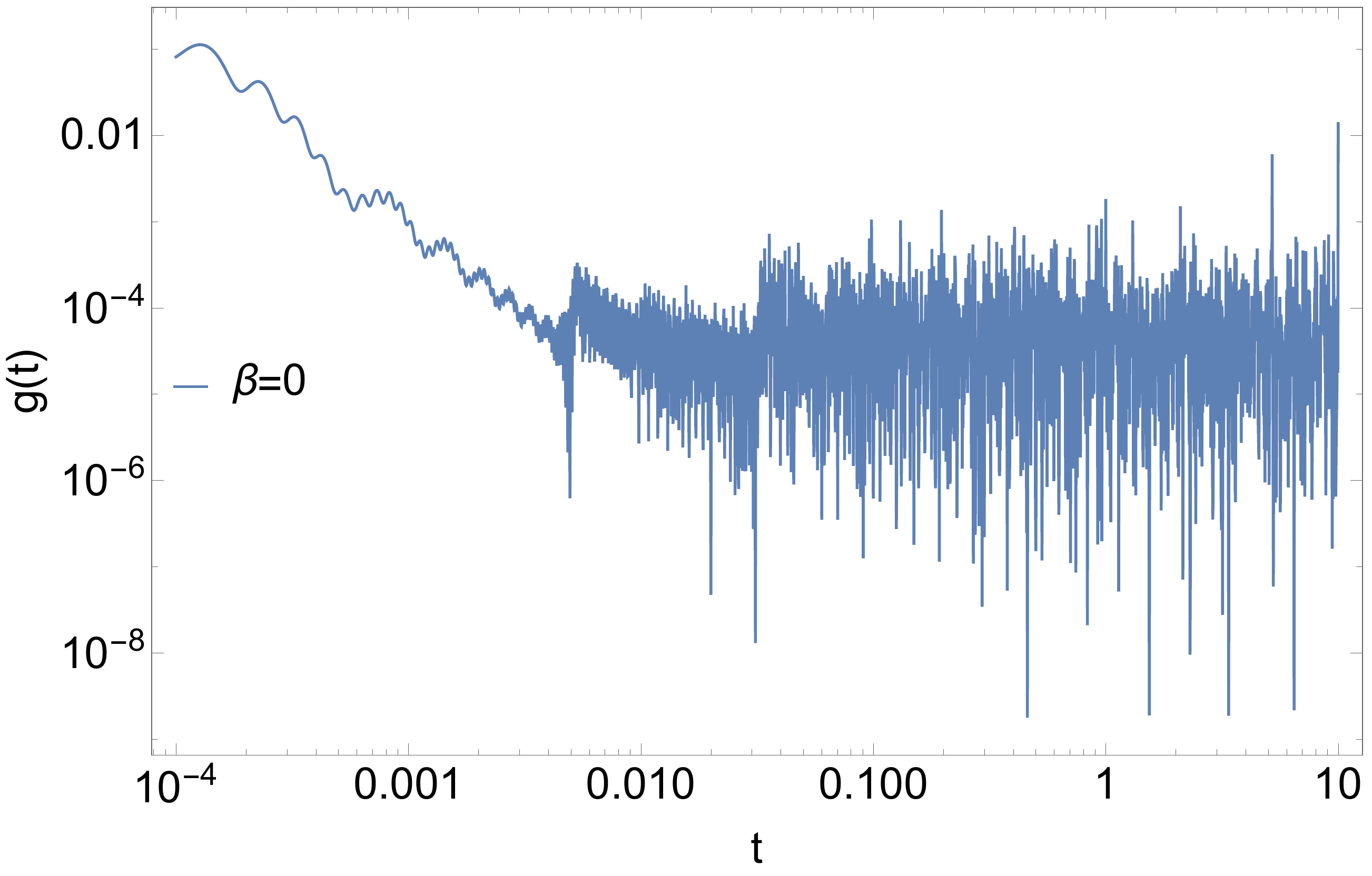}
    \end{subfigure}
    \hfill
    \begin{subfigure}{0.5\textwidth}
    \includegraphics[width=\textwidth]{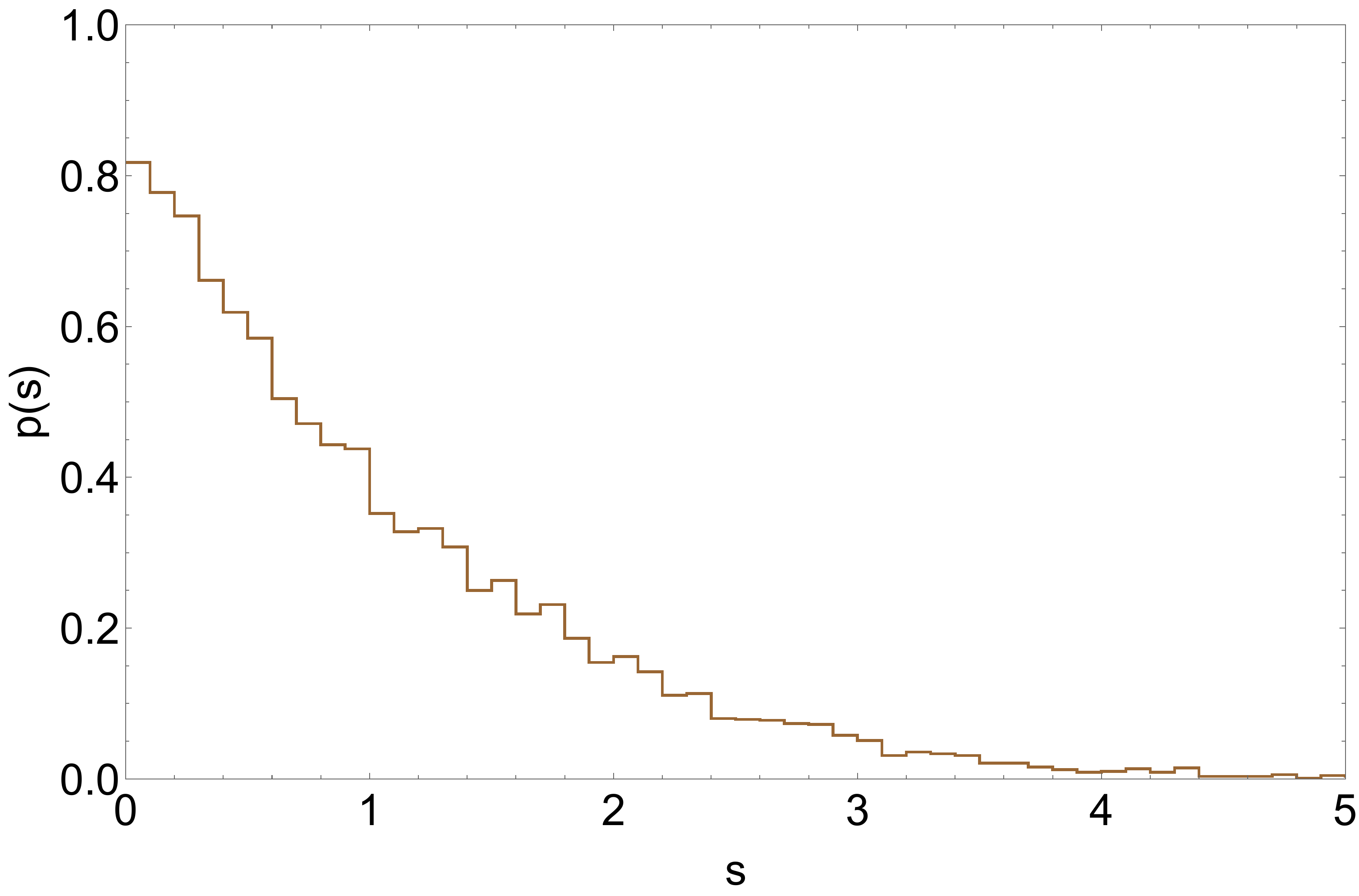}
    \end{subfigure}
    \caption{SFF and level spacing plot for rectangular billiard with $\alpha_1=1$, $\alpha_2=2\pi$, $n_{1 cut}=n_{2 cut}=100$.}
    \label{rec_bil_1}
\end{figure}

Rectangular billiard is an integrable system with energy levels given by,
\begin{equation}
    E_{n_1,n_2}=\alpha_1 n_1^2+\alpha_2 n_2^2.
\end{equation}
Here $n_1, n_2 \in \mathbb{N}$ and $\alpha_1, \alpha_2$ depend on the lengths of the sides. The corresponding level spacing is Poissonian for irrational ratio of $\alpha_1$ and $\alpha_2$ \cite{rectbilliards}. To calculate SFF we have used the following partition function,
\begin{equation*}
    Z(\beta,t)= \sum_{n_1=1}^{n_{1 cut}} \sum_{n_2=1}^{n_{2 cut}} e^{-(\beta-it)E_{n_1,n_2}}.
\end{equation*}
In Figure \ref{rec_bil_1} we have plotted the SFF with the corresponding level spacing distribution.

\FloatBarrier

\subsection{Square-Mod System}

\begin{figure}[h]
\begin{subfigure}{0.5\textwidth}
    \centering
    \includegraphics[width=\textwidth]{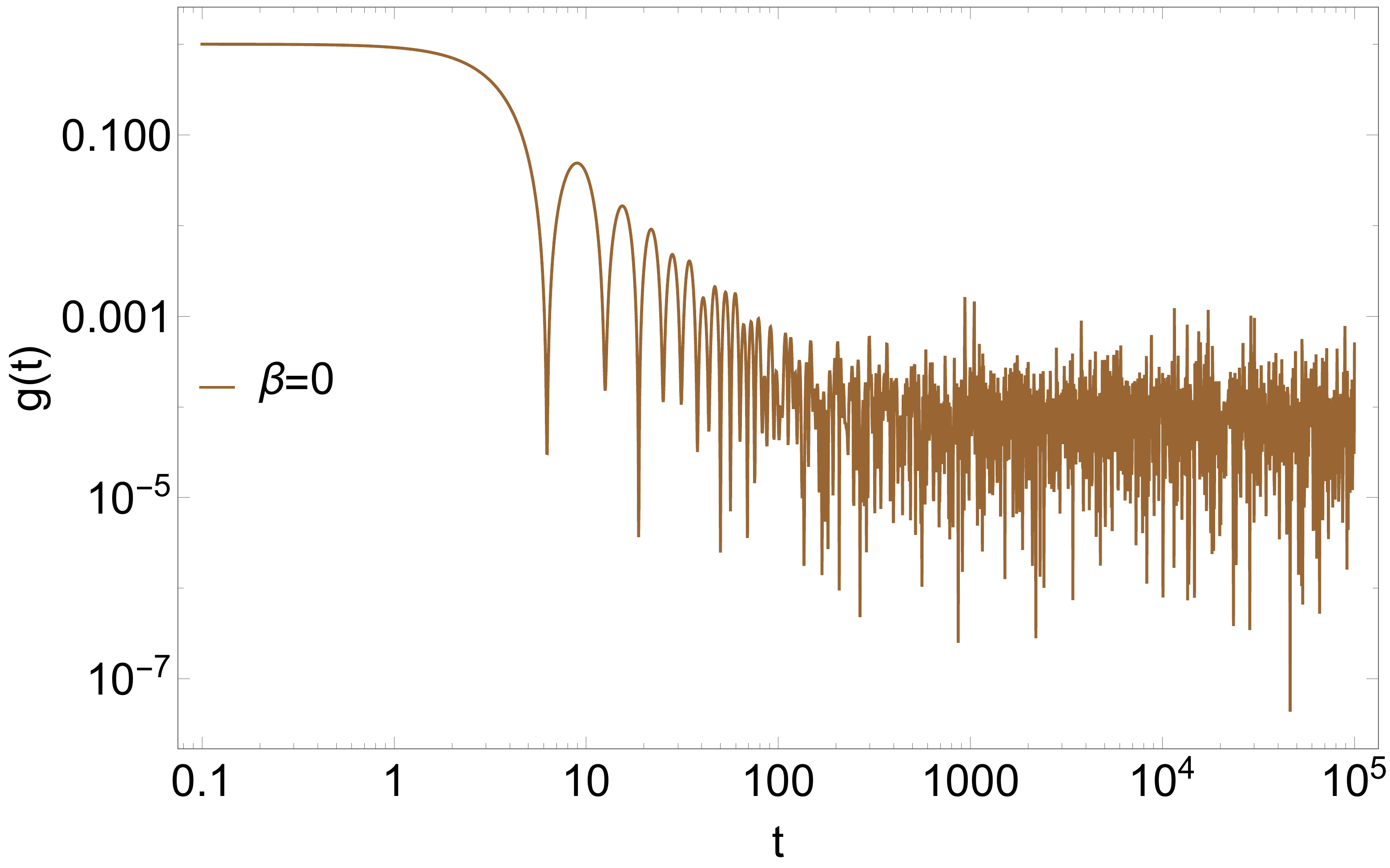}
    \end{subfigure}
    \hfill
    \begin{subfigure}{0.5\textwidth}
    \includegraphics[width=\textwidth]{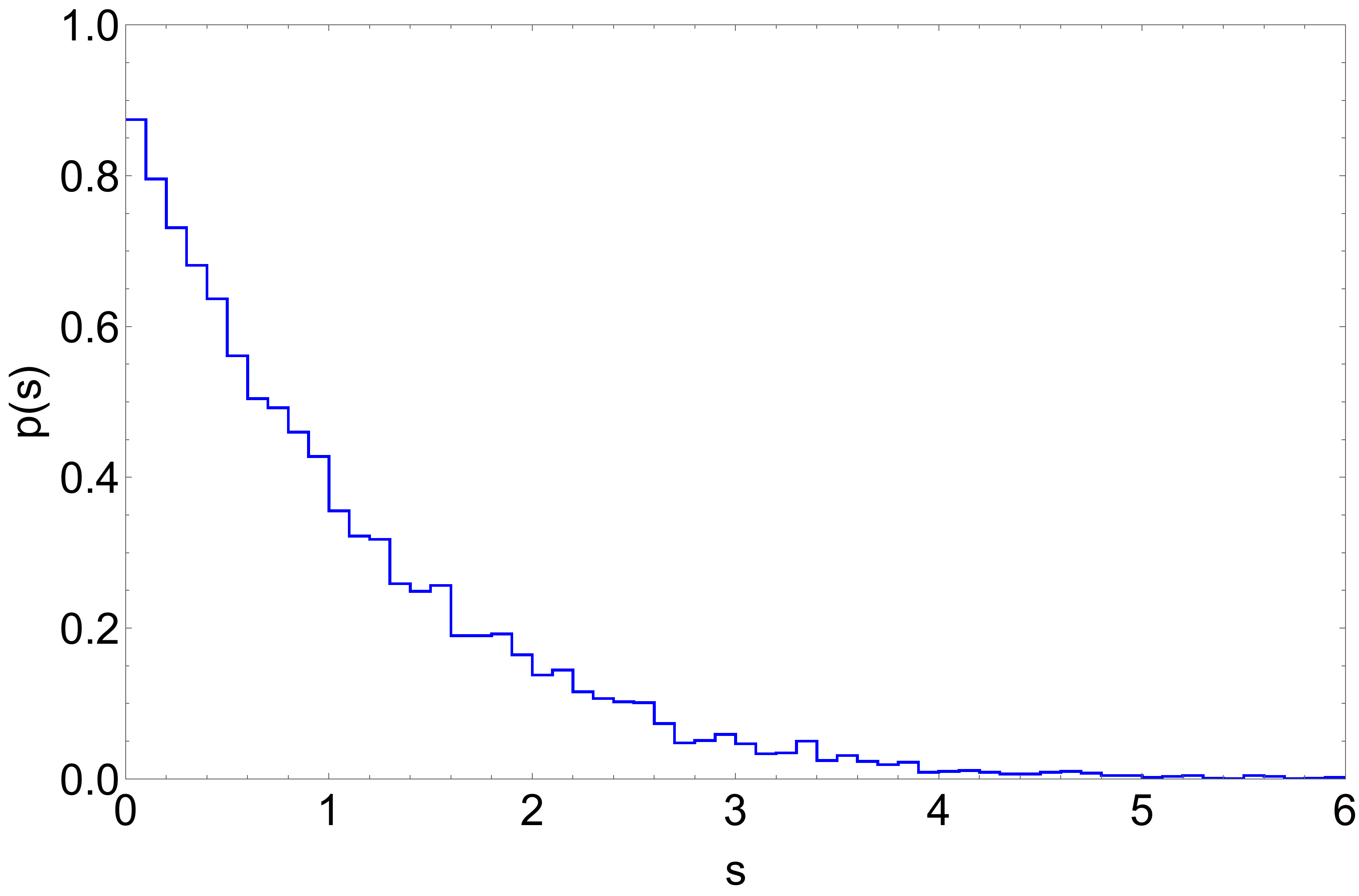}
    \end{subfigure}
    \caption{SFF and level spacing for square mod system with $j_{cut}=10000$ and $\alpha=\pi$.}
    \label{unknown_sff1}
\end{figure}

Consider a system with eigenvalue spectrum given by
\begin{equation}
    E_j=\alpha j^2 \ \text{mod}(1),
    \end{equation}
where $\alpha$ is an irrational number. This is known to be an integrable system that exhibits Poisson statistics \cite{rectbilliards}.
In Figure \ref{unknown_sff1}, we present the SFF and the corresponding level spacing distribution.

%


%
%
%
%


\subsection{SYK$_2$ Model}

In this section we consider the famous SYK model, but with only two (Majorana) fermion ``interactions''. The Hamiltonian is given by
\begin{equation}
    H=i\sum_{1\leq i_1 < i_2 \leq N}J_{i_1,i_2} \psi_{i_1}\psi_{i_2}.
\end{equation}
where the $J_{i_1,i_2}$'s are all-to-all couplings picked randomly from a Gaussian ensemble of mean $0$ and standard deviation $\frac{J}{\sqrt{N}}$. This is a quadratic system and therefore clearly integrable, what makes it interesting is the random coupling.

\begin{figure}[h]
\begin{subfigure}{0.5\textwidth}
    \centering
    \includegraphics[width=\textwidth]{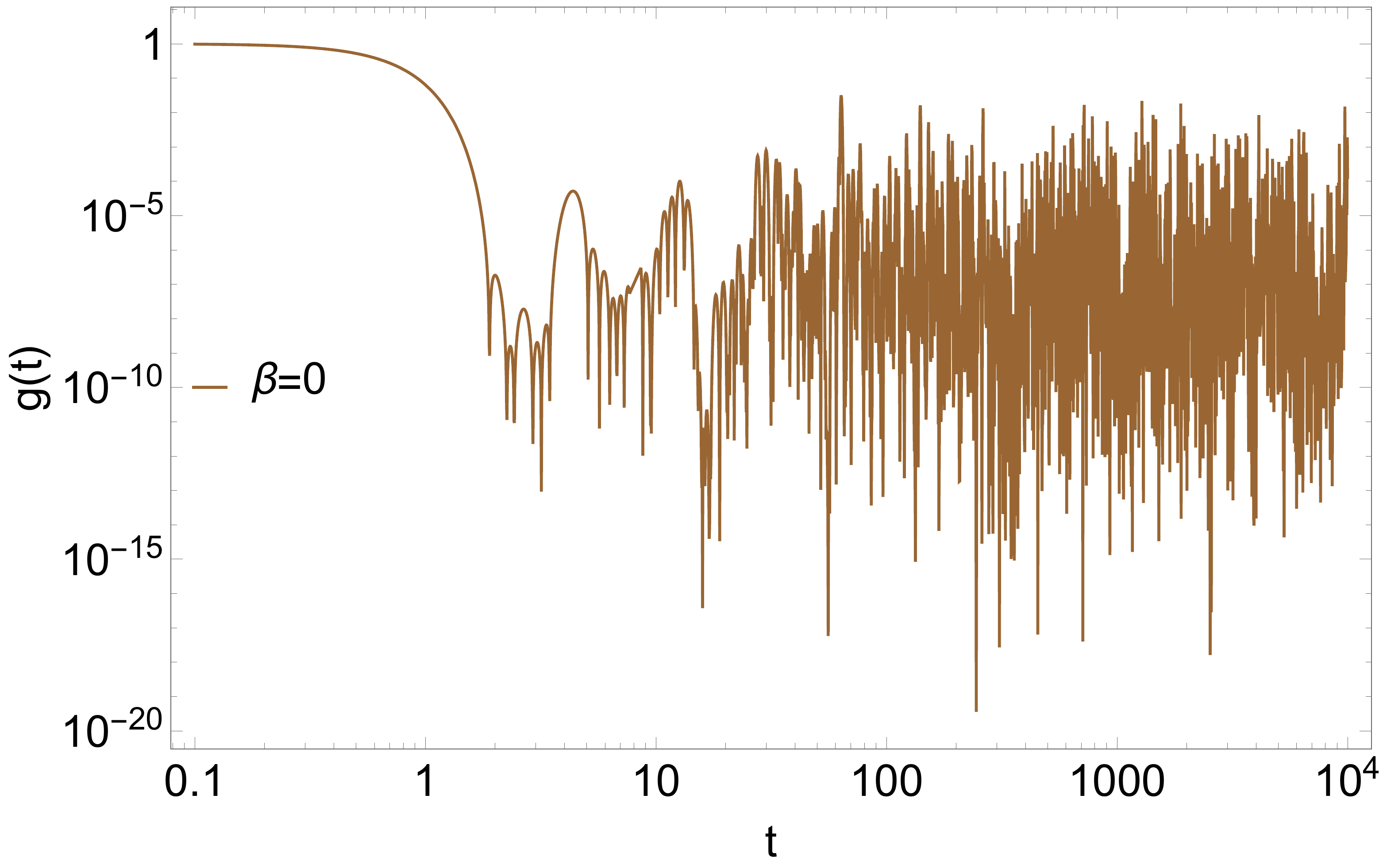}
    \end{subfigure}
    \hfill
    \begin{subfigure}{0.5\textwidth}
    \includegraphics[width=\textwidth]{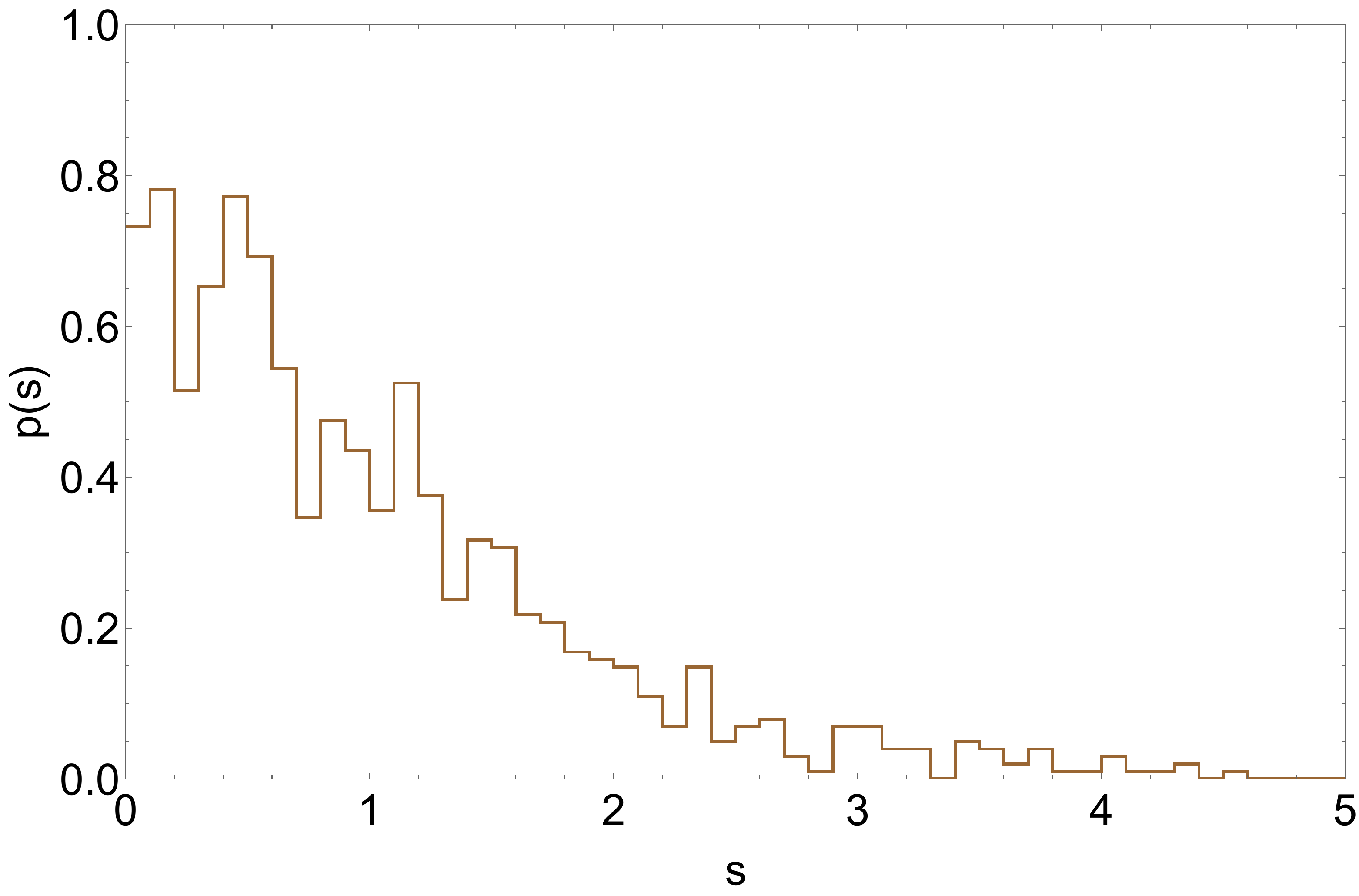}
    \end{subfigure}
    \caption{SFF and level spacing plot for SYK2 model with number of sites $N=24$, $J=1$ and without disorder averaging.}
    \label{syk_sff1}
\end{figure}

Figure \ref{syk_sff1} shows SFF for SYK$_2$ model for some representative choice of parameters, for a single realization of the random couplings. On the right panel, it also shows the level spacing for the same system. 
\FloatBarrier

As evident, the level spacing is consistent with Poisson, and the SFF does not show any clear DRP structure. So this system follows our narrative that (unlike the stretched horizon) integrable systems do not exhibit a ramp.

However, there is a small wrinkle in this story specifically in the case of SYK$_2$ model. An interesting point about the SYK$_2$ model is that if one does an average over the random couplings, a loosely ramp-like feature emerges -- this was noted in \cite{Jeff},\cite{chentema} and we illustrate this in Figure \ref{sff_rind3_avg}.  In other words, SYK$_2$ is an example of a system which exhibits a ramp {\em after} averaging over random couplings. It is not clear to us if there is a conceptual understanding of the connection between the ramp and the averaging over random couplings in general systems, but we have {\em not} seen this in the other systems we have considered in this section. 
the level spacing, for the same system. 
\vspace{0.11in}
\begin{figure}[h]
    \centering
    \includegraphics[width=.55\textwidth]{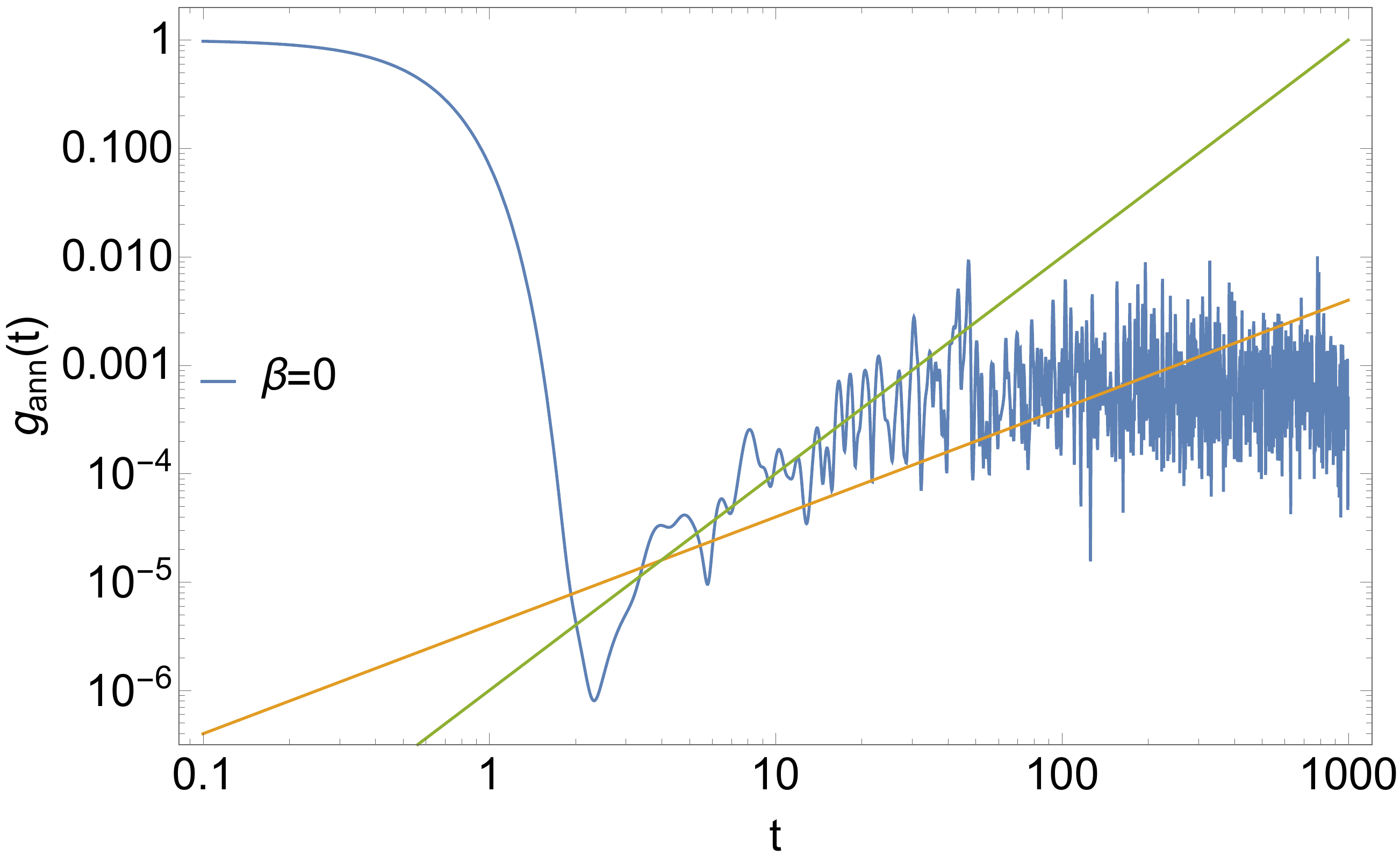}
    \caption{SYK2 model with number of sites $N=20$. The ramp is fit with linear lines of slope $1$ (yellow) and $2$ (green). Averaging is done over hundred randomly chosen $J$ from a normal distribution with mean $\mu=1$ and variance $\sigma=0.005$.}
    \label{sff_rind3_avg}
\end{figure}
\FloatBarrier
Doing an ensemble average over the parameters can lead to a clean dip and a plateau in many systems as simple as the SHO. We illustrate this for the SHO in one of the Appendices. But the emergence of a linear ramp is non-trivial and usually considered to be a signature of RMT behavior, and not expected in integrable systems. Indeed, one can check (see Figure \ref{sff_rind3_avg}) that the slope of the ramp in the SYK$_2$ model here is {\em not} consistent with $\sim 1$.  This is what makes our stretched horizon examples special.
%
%


\section{Discussion}

\subsection{Summary and Outlook}

Efforts to understand black holes as quantum objects in recent years have proceeded along two broadly (apparently) distinct paths. One of them adopts the fuzzball philosophy that microstates cap off at the horizon. The second philosophy (which for lack of a better name we will call the ``semi-classical approach'') proceeds by taking lessons from semi-classical gravity and entanglement entropy, seriously. 

Both these approaches have had their successes, but the full picture is yet to emerge. We have discussed some of the successes of the fuzzball program in the introductory section. The semi-classical approach has also had its successes, the most striking being the recent derivation of the Page curve via semi-classical islands \cite{Penington, Almheiri}. It is remarkable that quantum extremal surfaces (QES) \cite{Wall} which were proposed on quite independent grounds, have naturally lead to a semi-classical derivation of the Page curve. We emphasize that islands are a natural consequence of QES, and not an extra fixture to engineer the answer.

Of course, just as for fuzzballs, the information paradox is yet to be fully resolved even in the semi-classical approach. While the final Page curve in the island paradigm is compatible with unitarity, the detailed emergence of unitarity at each epoch of Hawking radiation has raised numerous questions \cite{PSSY} related to ensemble averaging and factorization. There is evidence that semi-classical gravity should be viewed as an ergodic proxy for a time average during each epoch of Hawking radiation \cite{Rozali, Hong, Vyshnav, Schlenker}. Finally, a satisfactory understanding of why the horizon is smooth (or a compelling argument why smoothness is not necessary) has not emerged from either the fuzzball program or the semi-classical approach.

Given this state of affairs, it is perhaps worthwhile trying to understand the successes of one approach, via the methods of the other. We suspect that both perspectives will have to be integrated, and neither simply ignored. This is not because of our wish to be reconciliatory, but because both approaches have had interesting results and possible successes, as well as profound challenges (as we mentioned above). A natural way to make progress might be to take the successes of one perspective, and see what one learns by trying to reproduce them from the other.

This paper should be viewed as an effort in this direction. Recent developments suggest that black hole horizons are fast scramblers \cite{Sekino, Butterfly, MSS} and that they exhibit features of random matrices  \cite{cotler}. Therefore, if the bulk description of black hole microstates is in terms of fuzzball ``geometries'' that cap off at the horizon, then we should be able to see hints of random matrices in these fuzzballs. We have taken a rudimentary step in this direction in this paper, by considering black hole geometries with a stretched horizon. We found that the scalar normal modes in such geometries have an SFF that exhibits the classic dip-ramp-plateau (DRP) structure with a clear linear ramp. This is surprising because typically a linear ramp is associated to level repulsion in the spectrum, which cannot exist for free modes in the stretched horizon geometry. Indeed, we explicitly checked that the spectrum does not exhibit conventional level repulsion, and yet it leads to the DRP structure. Typically, integrable Poisson distributed systems do not exhibit a ramp, except in some cases with random couplings where a ramp-like feature may appear after an ensemble average over the couplings. On the contrary, we found a linear ramp in our normal modes without any ensemble averaging. We traced its origins to the non-trivial level spacing distribution of normal modes as a function of the quantum numbers in the compact dimensions.  

These observations could be viewed as a suggestion that even in our extremely simple system (namely a black hole with a stretched horizon), there are hints of random matrices. This could be viewed as encouraging for the fuzzball program, but it also raises questions -- is there a way in which one can see level repulsion directly in a stretched horizon calculation using probe fields? This would be interesting, but considering the fact that RMT behavior really only need be exhibited by the full quantum gravitational black hole microstates, it is not obvious that such a calculation should exist. But we will present a proposal for such a calculation in the next subsection.

\subsection{Self-Interacting Probes?}

We start by observing that there is a stark distinction between SYK$_2$ and SYK$_{4}$ (or for that matter any SYK$_{q > 2}$) models. While the former has random couplings, and exhibits a ramp-like structure (with a slope $\nsim 1$) after an ensemble average, it does not exhibit level repulsion because it is a free theory. On the contrary, the interacting SYK models (those with $q > 2$) do exhibit level repulsion. This analogy is a suggestion that the natural thing to try in order to see level repulsion in our stretched horizon set up, is to make the probe scalar theory, interacting. There is another heuristic reason to suspect that probe self-interactions may be important. Quartic self-interacting scalar field has been noted to have physics with qualitative and quantitative parallels to the self-coupling of gravity responsible for the non-linear instability of AdS geometries \cite{Pallab1, Pallab2, Yang, CKOleg}. On the other hand, the non-linear instability of capped geometries \cite{Reall} has been argued to drive non-generic microstate geometries to generic microstate geometries \cite{Marolf}. In other words, non-linear instabilities triggered by gravitational interactions have a {\em generifying} effect on the microstate. This again is indicative of something that might be relevant when we are looking for RMT behavior, and suggests the possibility that a self-interacting field may be able to capture some of this physics\footnote{An interacting quantum field theory, at least in its high-lying part of the spectrum, is likely to exhibit RMT behavior, thanks to the Eigenstate Thermalization Hypothesis (ETH). It will be interesting to study what happens in the low-lying part of the spectrum, when such a theory is placed on a stretched horizon.}. Bosonic models which are closely related to SYK and AdS instability have appeared in \cite{Oleg1, Oleg2, Oleg3, Oleg4} and a connection between SYK models and AdS instability has been alluded to in \cite{Bala}. The relevance of the interacting scalar to the AdS instability problem was first noted in \cite{Pallab1}.

A third reason to suspect that interactions may be important is that they can produce level spacing that is non-perturbative in the interaction strength. It is known that the energy difference between the quasi-degenerate states of a double well potential go as $\sim \exp(-1/\lambda)$ \cite{ZinnJustin} where $\lambda$ is the quartic self-coupling. This is interesting because in a gravitational setting, the analogue is $\sim \exp(-1/G)$ which is the expected level-spacing between black hole microstates. In other words, without interactions it will be a challenge for the fuzzball paradigm  to find a mechanism that can reproduce the non-perturbative (in $G$) level spacing, if we are given a Planckian stretched horizon. 

The above arguments are heuristic, and so our speculations about interacting probe scalars (or fermions), needs more tests. Nonetheless, these arguments do make this possibility worth exploring --  the challenge is that this leaves us with the task of identifying the spectrum of an interacting quantum field theory, say $\phi^4$ theory, in the black hole background with a stretched horizon. The simplest setting to attempt this calculation may be in 1+1 d Rindler or Poincare horizon of AdS$_2$ (times possibly a compact space). None of these calculations are trivial, but their virtue is that unlike the task of determining the spectrum of fuzzballs from first principles, these are not entirely hopeless at this stage. 

\subsection{Finer Details and Open Questions}

In this final subsection we will make various comments that are relevant to the discussion in the main body of the paper, as well as point out some of the open questions. 
\begin{itemize}
\item We have observed that the slope of the ramp is $\sim 1$. There was some ambiguity in the precise location of the end point of the ramp structure, because the end of the ramp and the beginning of the plateau did not quite coincide -- in particular, there is a kink-like structure at the end of the ramp in some of our plots (see eg. Figure \ref{btz_many_realization} where it becomes very clear after an ensemble average). So a more precise statement about the slope was not very meaningful. It will be useful to understand the physics of the kink better.

\item More generally, do the ramps we find have anything to do with the chaos associated to the black hole horizon \cite{Butterfly}? If yes, in what way are they a signal of chaos? If not, what is the mechanism behind their robustness? The latter question is important because a linear ramp is widely used as a signal of RMT behavior.


\item As we noted previously, the compact dimensions were key for the emergence of the ramp. The precise nature of the compact dimensions did not seem to matter -- in the BTZ case, it was an angular direction associated to a radial coordinate, in the Rindler case it was simply a direct product space.
\item One of the observations that have emerged from our calculation is that even in a system without conventional level repulsion, we can have a linear ramp of slope $\sim 1$ on a log-log plot without any ensemble averaging. 
While it is known that the standard RMT classes exhibit a suitably defined linear ramp on a log-log plot\footnote{Scaled in terms of the Heisenberg time $t_H = 1/(\pi \Delta)$, where $\Delta$ is the mean level spacing, the behavior of the ramp takes the following forms in the most well-known random matrix ensembles:
\bea
{\rm GUE :} &=& \tau \\ 
{\rm GOE :} &=& 2 \tau + \log \tau \ {\rm correction} \\
{\rm GSE:} &=& 2\tau + \tau \log \tau \ {\rm correction} 
\eea
where $\tau = t/t_H$. See \cite{Altland, Haake} for related discussions. We thank Julian Sonner for explanations on this.}, it is not clear to us if the converse statement is true. The microscopic origin of the ramp can contain extra information, and in order to pin down a linear ramp as due to random matrix behavior, we need more detailed information. 
What  is remarkable is that we seem to be finding a linear ramp of slope $\sim1$ in our plots, across multiple systems. This suggests that finding linear ramps without RMT spectra can be generic in some suitable sense. Considering the fact that generic integrable systems do not have ramps, and the fact that RMT systems do, it will clearly be useful to understand what aspect of the spectrum is being captured by the linear ramp in our examples. 


\item It should also be noted that the ramp that we found, did not require an ensemble average.  We emphasize this, because as we discussed in the text, we know of one example where an ensemble average over random ``couplings'' in an integrable theory leads to a ramp. This is the SYK$_2$ model that we discussed in the previous section. To repeat, there are two facts that distinguish SYK$_2$ from our stretched horizon modes. The first is that without an average, there is no ramp there. Secondly, even after the average, the ramp that emerges is {\em not} linear of slope $\sim1$. An ensemble average over parameters can lead to a dip and a plateau even in the Simple Harmonic Oscillator (see Appendix), but a linear ramp of slope $~1$ is less trivial.

\item 
It is interesting to compare the behavior of the spectrum at a finer level. For example, following \cite{cotler}, the RMT dip time is expected to behave as $t_{\rm dip} \sim \sqrt{L}$, where $L$ is the dimension of the Hilbert space = rank of the matrix. While it is difficult to conclude a precise dependence, our numerical data do exhibit a growing behaviour of the dip time, with the dimension of the Hilbert space. For us, the cut-offs, $n_{cut}$ and $J_{cut}$ set the effective dimension of the Hilbert space. It will be interesting to make it more precise. 

\item Our results show that the normal modes of the stretched horizon have features that straddle the boundary between integrable and chaotic behavior. Is it possible to identify a non-black hole system where the SFF shows a ramp, but the spectrum is {\em not} one  of the conventional RMT classes? As already noted, one feature of our plots is that in some of them (see eg. Figure \ref{btz_many_realization}) after the ramp, there seems to be a small kink before the curve stabilizes to the plateau. It will be interesting to understand the spectral origin of this feature, in a non-black hole spectrum. 

\item 
Note that the basic ingredients of our calculations are, at a technical level, very similar to the brick wall approach of 't Hooft in \cite{tHooft:1984kcu}. This approach provides a natural mechanism of producing a large number of localized states near the horizon and therefore can produce a horizon-area worth of entropy. 
Our calculation further illustrates that a brick wall type scenario may be capable of capturing some aspects of the fine-grained information of the quantum black hole. This observation further emphasizes the question: What class of quantum fine-grained quantities are accessible within a semi-classical description of gravity? Answering this is beyond the scope of this work, but we would like to explore this issue in future.

\item A paper that has discussed ideas of a somewhat similar character to ours is \cite{Craps}. They considered two point functions in the weakly coupled D1-D5 system as a proxy for SFF, and found that there is a dip-ramp-plateau structure. There are quite a few  differences between their calculation and ours, we will emphasize a few. Their calculation was in the boundary theory (in the free limit), we are doing our calculation in the bulk. They work with a two point function, not the SFF\footnote{This, for example, leads to differences in the plateau height. In typical SFF calculations, the plateau height is exponential in the entropy of the system. This is not the case in \cite{Craps}. Note for example that the plots of their 2-point functions are on linear-log axes, unlike the log-log axes typically used in the SFF.}. The state in which the calculation was done was typical in the sense that the twist distribution of the (Ramond ground) state was chosen to be in the grand canonical ensemble. In order to see the ramp (which was logarithmic), they do a running time average. We see the ramp in the SFF without any time averages.

\item The existence of the ramp in the semi-classical (quasi-Euclidean) gravity approach is understood in terms of replica wormholes \cite{Saad}. This leads to an averaged ramp where fluctuations are suppressed. A holy grail is to see the origin of the ramp, but {\em without} averaging and {\em with} fluctuations from a bulk quantum gravity calculation. Our calculation, while certainly not in full quantum gravity, was meant to be a toy model for it -- and we find both the ramp and the fluctuations.

\item What do these results say about the fuzzball program? That a {\em quantum} fuzzball microstate is a regular configuration that caps off at the horizon, is not an operationally well-defined statement yet, in our view\footnote{Note that currently the concrete successes of the fuzzball program have been in the setting of zero temperature black holes and classical gravity. See some recent discussions \cite{Witten} which suggest that in the setting of quantum field theory and perturbatively weak gravity at finite temperature, one can associate density matrices and entropies to the exterior of a black hole, but not pure states. This is a von Neumann algebra of Type II. According to the fuzzball program, it seems that non-perturbatively in $N$, one should obtain a type I algebra ``outside the horizon''. The challenge is of course to make sense of this statement (and bulk causality in general) at finite $N$.}. However we find the heuristic picture inspiring. The fact that this heuristic picture lead to some expected  structures in the SFF is striking, and in that sense our result can be viewed as encouraging for the fuzzball program. But a counter-argument is that perhaps the ramp is not a signature of RMT behavior as is widely believed, and may be signaling other features of the spectrum. This would be surprising, but this is a logical possibility and if this is the case, more work is needed to identify precisely what those features are. Finally, progress on sharpening the notion of a {\em quantum} fuzzball at finite temperature will be useful, given the absence of explicit bulk solutions.


\item Our paper has been largely phenomenological. We observed a suggestive and robust linear ramp in stretched horizons with a compact extra dimension, but we did not provide a deep understanding of its conceptual origin. 
But it is evident that the quasi-degeneracy of levels in the $J$-direction is crucial for what we see. A deeper understanding of this may be a useful step towards clarifying the connection between microstate geometries and the nature of the horizon.

\end{itemize}





\section{Acknowledgments}

We thank Souvik Banerjee, Sumilan Banerjee, Oleg Evnin, Sumit Garg, Emil Martinec, Pradipta Pathak and Julian Sonner for discussions and Sthitadhi Roy for a code snippet that helped us double check some of our results.

\appendix

\section{Computing Argument of $P/Q$}

In this an the next Appendix, we demonstrate some simple facts which are crucial for the calculations of the BTZ normal modes. From equations \eqref{PQ1} and \eqref{PQ2} with $\nu=1$ we have,
\begin{align}\label{Z}
    \frac{P}{Q} &=\frac{\Gamma(i \omega)e^{\pi \omega}\left(e^{J\pi}+e^{\pi(-\omega+i)}\right) \Gamma\left(\frac{1}{2}(-i\omega-i J)\right) \Gamma\left(\frac{1}{2}(2-i\omega-i J)\right)}{\Gamma(-i \omega)\left(e^{J\pi}+e^{\pi(\omega+i)}\right) \Gamma\left(\frac{1}{2}(i\omega-i J)\right) \Gamma\left(\frac{1}{2}(2+i\omega-i J)\right)} \nonumber \\
    &= \frac{\Gamma(i\omega)}{\overline{\Gamma(i\omega)}} \left(\frac{\Gamma\left(-\frac{i}{2}(\omega+J)\right)}{\Gamma\left(\frac{i}{2}(\omega-J)\right)}\right)^2 \underbrace{ \left(\frac{e^{\pi(J+\omega)}-1}{e^{J\pi}-e^{\omega\pi}}\frac{(J+\omega)}{(J-\omega)}\right)}_{\equiv A_1}.
\end{align}
Equating arguments of both sides,
\begin{align}
    \text{Arg}\left(\frac{P}{Q}\right) &=\text{Arg}\left( \frac{\Gamma(i\omega)}{\overline{\Gamma(i\omega)}} \left(\frac{\Gamma\left(-\frac{i}{2}(\omega+J)\right)}{\Gamma\left(\frac{i}{2}(\omega-J)\right)}\right)^2\right)+\text{Arg}(A_1)\nonumber \\
    &=2 \text{Arg}\left(\Gamma(i\omega)\right)+ \text{Arg}\left(\frac{\Gamma\left(-\frac{i}{2}(\omega+J)\right)}{\Gamma\left(\frac{i}{2}(\omega-J)\right)}\right)^2+\text{Arg}(A_1).\nonumber
\end{align}
Now for all positive $\omega$ and both positive and negative $J$, $A_1$ is always positive so it has zero contribution in the argument. So,
\begin{align}
    \text{Arg}\left(\frac{P}{Q}\right) &=2 \text{Arg}\left(\Gamma(i\omega)\right)+ \text{Arg}\left(\frac{\Gamma\left(-\frac{i}{2}(\omega+J)\right)}{\Gamma\left(\frac{i}{2}(\omega-J)\right)}\right)^2 \nonumber \\
    &=2 \text{Arg}\left(\Gamma(i\omega)\right)+2\text{Arg}\left(\Gamma\left(-\frac{i}{2}(\omega+J) \right) \right)-2\text{Arg}\left(\Gamma\left(\frac{i}{2}(\omega-J) \right) \right)\nonumber \\
     &=2 \text{Arg}\left(\Gamma(i\omega)\right)+2\text{Arg}\left(\Gamma\left(-\frac{i}{2}(\omega+J) \right) \right)+2\text{Arg}\left(\Gamma\left(\frac{i}{2}(J-\omega) \right) \right).
\end{align}

\section{Showing $|P|=|Q|$}

For any $k \in \mathbb{R}$ we have the following identities
\begin{itemize}
    \item $\Gamma(-i k)=\overline{\Gamma(i k)}$,
    \item $\Gamma(1+i k)=ik\Gamma(i k)$,
    \item $\Gamma(ik)\Gamma(1-ik)=\frac{\pi}{\sin{(i\pi k)}}=\frac{i\pi}{\sinh{(\pi k)}}$.   
\end{itemize}
Using the above identities one can easily check that
\bea
    |\Gamma(i\omega)|&=&|\Gamma(-i\omega)|, \\ 
    \Gamma(i k) \Gamma(-i k)&=&\frac{\pi}{k \sinh{(\pi k)}} \\
    \bigg|\Gamma\left(\frac{-i}{2}(\omega+J)\right)\bigg|^2 &=&\frac{2\pi}{(\omega+J)\sinh{(\frac{\pi(\omega+J)}{2})}}, \\
   \bigg|\Gamma\left(\frac{i}{2}(\omega-J)\right)\bigg|^2 &=&\frac{2\pi}{(\omega-J)\sinh{(\frac{\pi(\omega-J)}{2})}} 
\eea
Now, from equation \eqref{Z} we have
\begin{equation}\label{modpq}
    \frac{|P|}{|Q|} =\frac{|\Gamma(i\omega)|}{|\overline{\Gamma(i\omega)}|} \left(\frac{|\Gamma\left(-\frac{i}{2}(\omega+J)\right)|}{|\Gamma\left(\frac{i}{2}(\omega-J)\right)|}\right)^2 \frac{\left(e^{\pi(J+\omega)}-1\right)}{\left(e^{J\pi}-e^{\omega\pi}\right)}\frac{(J+\omega)}{(J-\omega)}
\end{equation}
Putting together all the above identities it follows immediately that $\frac{|P|}{|Q|}=1$.

\section{Ensemble Averaged SFF of the Harmonic Oscillator}

\begin{figure}[h]
\begin{subfigure}{0.5\textwidth}
    \centering
    \includegraphics[width=\textwidth]{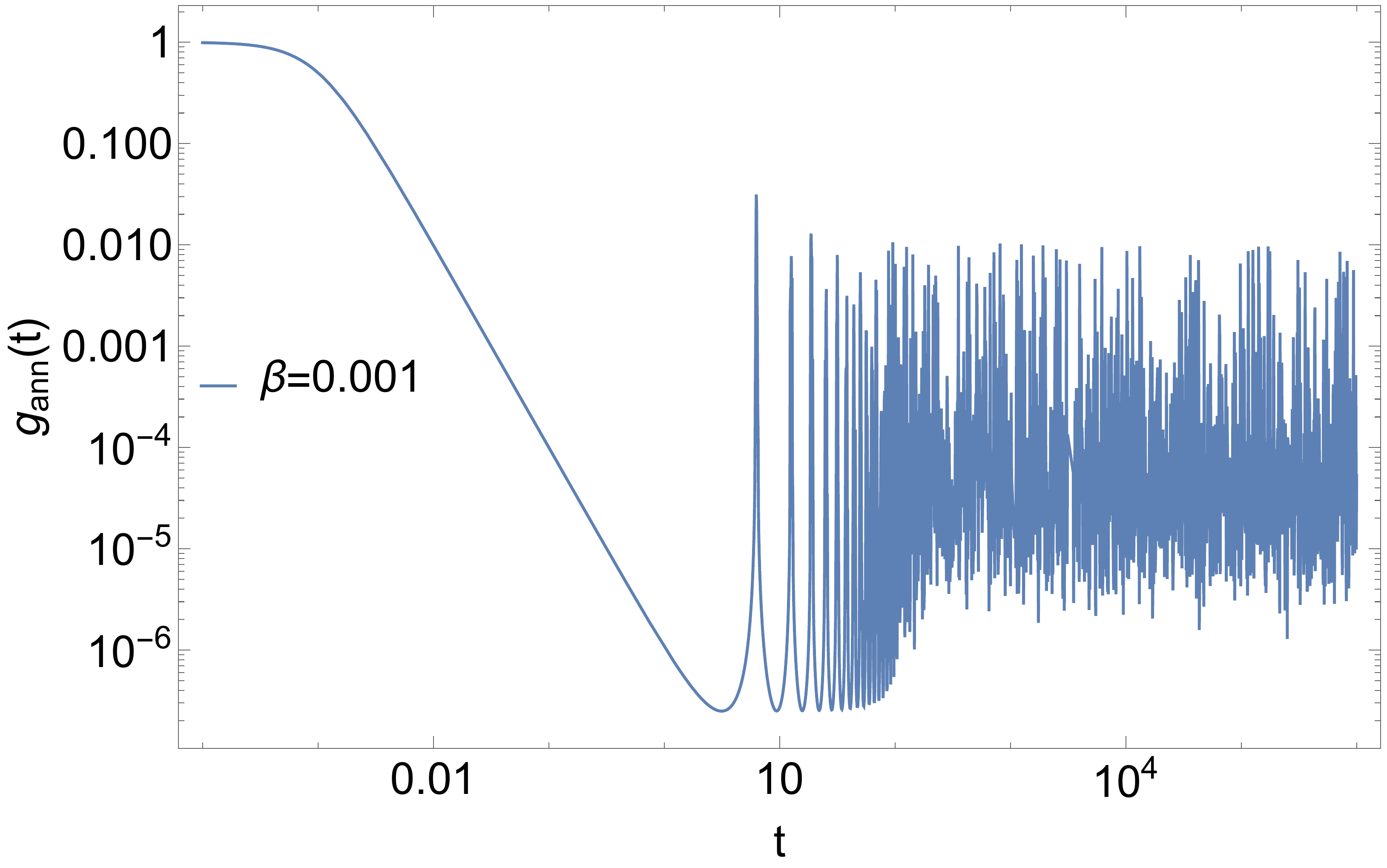}
    \end{subfigure}
    \hfill
    \begin{subfigure}{0.5\textwidth}
    \includegraphics[width=\textwidth]{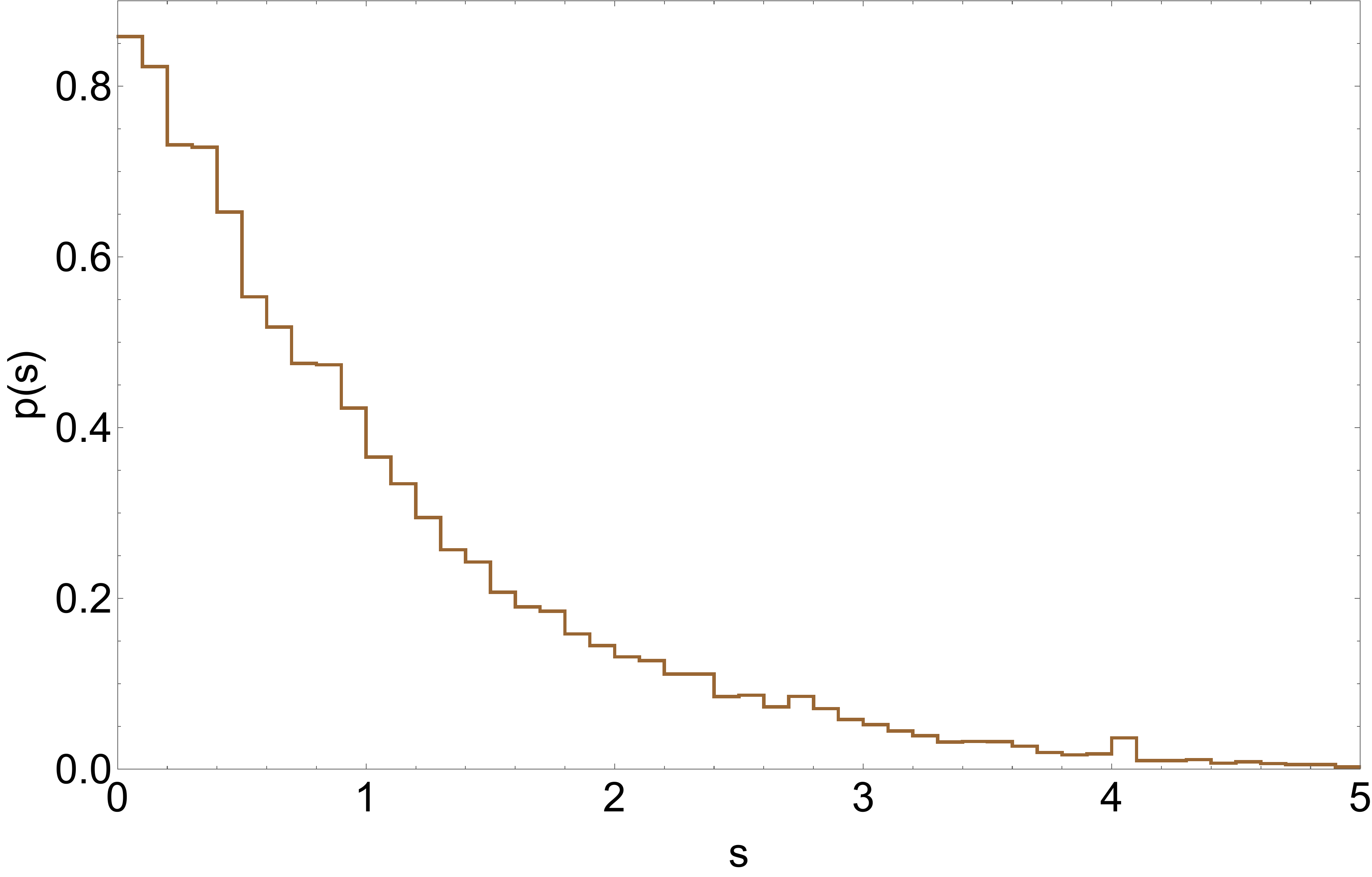}
    \end{subfigure}
    \caption{ Annealed SFF and its level spacing for SHO with $n_{cut}=300$, $\beta=0.001$. Averaging is done over hundred randomly chosen $\omega$ from a normal distribution with mean $\mu=1$ and variance $\sigma=0.01$.}
    \label{sho_sff2}
\end{figure}

Here we will illustrate that ensemble averaging can bring out a dip and a plateau even in the simplest integrable system, namely the SHO. The explicit form of the SFF in this case is trivial to compute:
\bea
g(\beta,t) = \frac{\cosh{\beta \omega}-1}{\cosh{\beta \omega}-\cos{\omega t}}
\eea
In the plots below, we have introduced randomness in the system by choosing different $\omega$'s randomly from a normal distribution of give mean and standard deviation. The corresponding SFF is shown  in Figure \ref{sho_sff2}.

One point of these plots is that one can get a dip and a plateau quite easily in integrable systems by working with an ensemble. The SHO example also illustrates that if you consider an ensemble of systems, the level spacing is always of a Poisson-like form\footnote{Note that for a single realization of an SHO,  the level spacing distribution would have been a delta function.}. This is generic and true even for an ensemble of RMT systems -- there is no mechanism to implement level repulsion {\em between} individual systems.

\end{document}